\documentclass[pdflatex,sn-nature]{sn-jnl}
\usepackage{graphicx}%
\usepackage{multirow}%
\usepackage{amsmath,amssymb,amsfonts}%
\usepackage{amsthm}%
\usepackage{mathrsfs}%
\usepackage[title]{appendix}%
\usepackage{xcolor}%
\usepackage{textcomp}%
\usepackage{manyfoot}%
\usepackage{booktabs}%
\usepackage{algorithm}%
\usepackage{algorithmicx}%
\usepackage{algpseudocode}%
\usepackage{listings}%
\theoremstyle{thmstyleone}%
\theoremstyle{thmstyletwo}%
\usepackage{soul}
\usepackage{tabularx}
\theoremstyle{thmstylethree}%
\usepackage{lmodern}
\usepackage{anyfontsize}

\usepackage{caption}
\usepackage{makecell}
\usepackage{adjustbox}
\usepackage{rotating}
\usepackage{siunitx}

\begin{document}
\title[Article Title]{A $6.37\,{\rm Hz}$ quasi-periodic oscillating photospheric emission in GRB~240825A}
\author[1]{\fnm{Guo-Yu} \sur{Li}}\email{guoyu.li@st.gxu.edu.cn}
\author*[1]{\fnm{Da-Bin} \sur{Lin}}\email{lindabin@gxu.edu.cn}
\author[1]{\fnm{Zhi-Lin} \sur{Chen}}\email{chenzl@st.gxu.edu.cn}
\author[2]{\fnm{Bao-Quan} \sur{Huang}\email{huangbaoquan@xmu.edu.cn}}
\author[2]{\fnm{Tong}\sur{Liu}\email{tongliu@xmu.edu.cn}}
\author[1]{
\fnm{En-Wei}\sur{Liang}\email{lew@gxu.edu.cn} }

\affil[1]{Guangxi Key Laboratory for Relativistic Astrophysics, School of Physical Science and Technology, Guangxi University, Nanning 530004, People's Republic of China}
\affil[2]{Department of Astronomy, Xiamen University, Xiamen, Fujian 361005, China}


\abstract{
Using data from Swift-BAT and Fermi-GBM,
we report the first detection of a high-confidence quasi-periodic oscillation (QPO) in the thermal emission of gamma-ray burst GRB~240825A.
The spectral analysis of the burst reveals two radiation components,
including a thermal emission dominant in $100 \text{--} 300\,\mathrm{keV}$ and a non-thermal emission spanning a wide energy range.
During the interval $2.07 \text{--} 3.25\,\mathrm{s}$ post-trigger, 
a strong QPO signal at $6.37\,\mathrm{Hz}$ ($\gtrsim 5\sigma$ confidence) is identified in the $100 \text{--} 300\,\mathrm{keV}$ thermal-dominated band.
The variability analysis of the non-thermal component ($15 \text{--} 30\,\mathrm{keV}$) uncovered a $0.67\,\mathrm{Hz}$ QPO, 
consistent with light-curve modeling using periodic fast-rise exponential decay pulses.
The strong QPO in the photospheric emission directly indicates a quasi-periodic oscillating jet.
Together with the non-thermal emission variability,
we show that this QPO can be explained in terms of a helical structure in the jet, where the viewing angle to the dominant emission region in the jet undergoes periodic changes.
}

\maketitle
Gamma-ray bursts (GRBs), featured with powerful bursts of $\gamma$-rays,
are the most luminous electromagnetic explosions in the universe\cite{2015PhR...561....1K}.
They are typically classified into short- and long-duration GRBs,
where short-duration GRBs are considered to originate from the mergers of compact objects\cite{2014ARA&A..52...43B,2017ApJ...848L..13A} and long-duration GRBs are linked to the collapses of massive stars\cite{2006ARA&A..44..507W}.
In addition, they are generally believed to be associated with ultrarelativistic jets that are launched from the central engines: a massive millisecond magnetar\cite{1992Natur.357..472U,1998A&A...333L..87D,2001ApJ...552L..35Z} or a stellar black hole (BH) surrounded by a hyperaccretion disk\cite{1993ApJ...405..273W,1999ApJ...518..356P}.
The internal energy dissipation of ultrarelativistic jets leads to the production of prompt $\gamma$-ray emission,
with prevalent emission models including photospheric models\cite{1994MNRAS.270..480T,2000ApJ...530..292M,2009ApJ...703.1696Z},
internal shock models\cite{1994ApJ...430L..93R,2020MNRAS.493.5218S},
and magnetic reconnection models\cite{2011ApJ...726...90Z}.
The prompt $\gamma$-rays are well known to be highly variable\cite{2018pgrb.book.....Z}.
The quasi-periodic oscillations (QPOs),
which stemmed from the central engine of GRB,
are particularly intriguing phenomena.
However, searches for QPOs in GRBs generally yielded results
with low significance levels or negative results.

In this paper, we report the detection of a strong oscillation at a frequency $6.37\,{\rm Hz}$ in the photospheric emission of GRB~240825A,
with a high confidence level ($\gtrsim 5\sigma$).
This is the first QPO that was discovered in the thermal component of GRBs.

\begin{figure}
    \centering
    \includegraphics[width=1.0\textwidth]{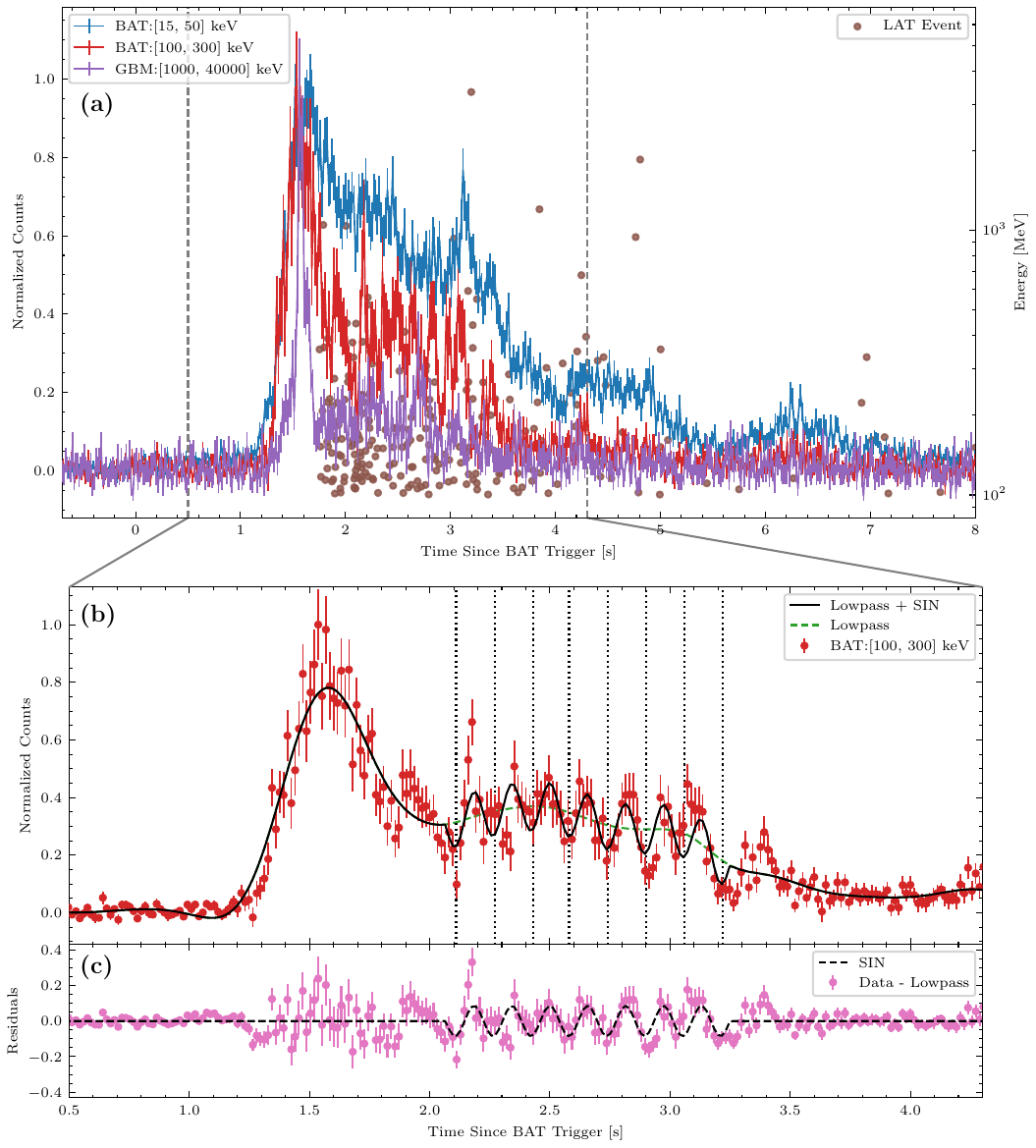}
\caption{Prompt emission of GRB~240825A detected by Swift-BAT and Fermi-GBM/LAT.
(\textbf{a}).  The light-curves of GRB~240825A at $15\text{--}50\,{\rm keV}$ (blue line), $100\text{--}300\,{\rm keV}$ (red line), and $1\text{--}40\,{\rm MeV}$ (violet line) are shown with time bin $0.016\,{\rm s}$.
The high-energy photon events detected by the Fermi-LAT are plotted individually with brown dots.
The light-curve of $100\text{--}300\,{\rm keV}$ presents a identifiable quasi-periodic modulation during $\mathord{\sim}2.07 \text{--} 3.25\,{\rm s}$.
(\textbf{b}). The zoom-in details of the $100 \text{--} 300\,{\rm keV}$ light-curve in $0.5 \text{--} 4.3\,{\rm s}$ are shown.
For clarity, a smooth low-pass curve (green dashed-line) modulated by a sinusoidal signal (with frequency of $6.33\,\mathrm{Hz}$) is plotted with a black curve to help indicate the QPOs in the $100 \text{--} 300\,{\rm keV}$ light-curve (See Methods for details).
(\textbf{c}). ``Residuals'' is obtained by subtracting the low-pass curve from the data. The quasi-periodic modulation is fairly obvious during $\sim 2.07\text{--}3.25\,\mathrm{s}$ in ``Residuals''.}\label{Fig_multi_lc}
\end{figure}

\begin{figure}
    \centering
\includegraphics[width=0.48\textwidth]{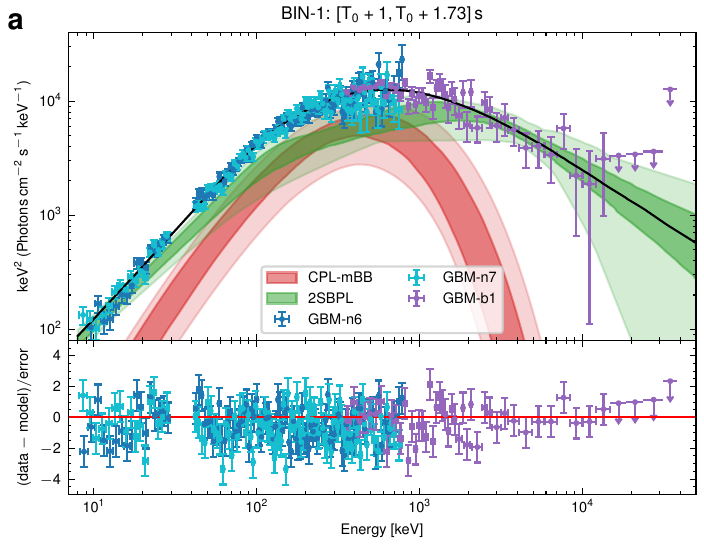}
\includegraphics[width=0.48\textwidth]{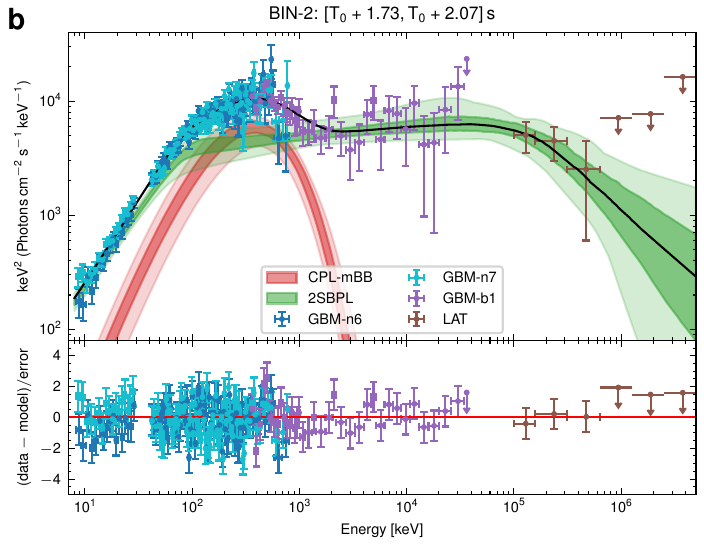}
\includegraphics[width=0.48\textwidth]{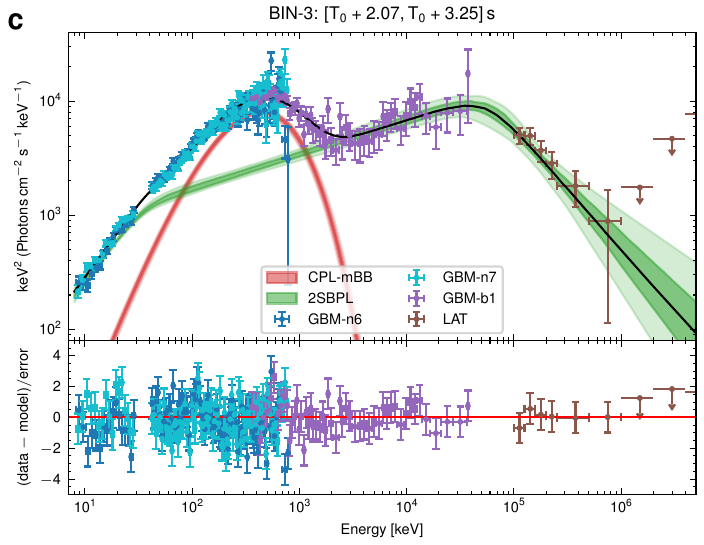}
\includegraphics[width=0.48\textwidth]{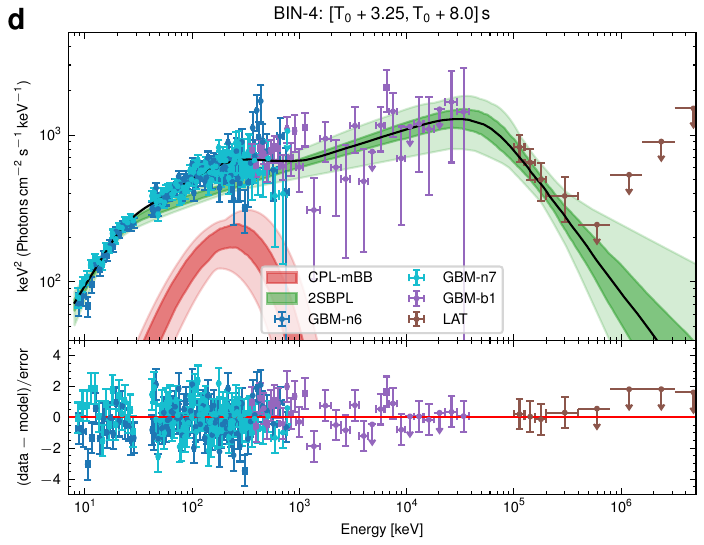}
\caption{Our spectral analysis on the time intervals of $1.0 \text{--} 1.73\,{\rm s}$, $1.73 \text{--} 2.07\,{\rm s}$, $2.07 \text{--} 3.25\,{\rm s}$, and $3.25 \text{--} 8.0\,{\rm s}$, respectively.
Two radiation components are identified:
a thermal emission at around $300\,{\rm keV}$ and a non-thermal emission spanning a wide energy range.
The thermal component dominates the $100 \text{--} 300\,{\rm keV}$ flux in $2.07 \text{--} 3.25\,{\rm s}$ and shows an obvious quasi-periodic modulation on its light-curve.
Since the thermal component is from the photosphere of the jet,
the QPO modulation feature directly signifies the QPO behavior of the GRB's jet.
For display, data points have been rebinned at $3\sigma$ significance
or grouped in sets of 6 bins/3 bins for GBM/LAT data.
Error bars indicate the $1\sigma$ uncertainty on data points, 
and arrows indicate upper limit at 95\% confidence level. }\label{Fig_spectrum}
\end{figure}

On 2024 August 25 at 15:52:59.84 UTC (hereafter $T_{0}$),
the Neil Gehrels Swift Observatory's (Swift) Burst Alert Telescope (BAT) triggered and located GRB~240825A\cite{2024GCN.37274....1G}.
The burst was identified simultaneously by the Gamma-ray Burst Monitor (GBM) aboard the Fermi Gamma-ray Space Telescope at $\mathord{\sim} T_{0} + 0.25\,{\rm s}$\cite{2024GCN.37301....1S}.
The host galaxy of GRB~240825A was reported with redshift $z=0.659$\cite{2024GCN.37293....1M}.
The main emission phase of this burst begins at $\mathord{\sim} T_{0} + 1.2\,{\rm s}$
and lasts around $10\,{\rm s}$.
In Fig.~\ref{Fig_multi_lc}\textbf{a}, we show the light-curves at three different energy bands,
including $15\text{--}50\,{\rm keV}$, $100\text{--}300\,{\rm keV}$, and $1\text{--}40\,{\rm MeV}$.
Interestingly, the $100 \text{--} 300\,{\rm keV}$ light-curve seems to suffer from transient quasi-periodic modulation during $\mathord{\sim} 2.07 \text{--} 3.25\,{\rm s}$.
However, this QPO modulation feature is not present in the light-curves of $15\text{--}50\,{\rm keV}$ and $1\text{--}40\,{\rm MeV}$ (see Extended Data Fig.~\ref{Fig_wwz_compare} of Methods).
It signifies that $\rm 100 \text{--} 300\,keV$ emission
during $\mathord{\sim} 2.07 \text{--} 3.25\,{\rm s}$
has a very different origin
from that of the other two energy bands (see Methods).
Thus, we perform the spectral analysis on the phases of $1 \text{--} 1.73\,{\rm s}$, $1.73 \text{--} 2.07\,{\rm s}$, $2.07 \text{--} 3.25\,{\rm s}$, and $3.25 \text{--} 8\,{\rm s}$ (see Spectral Analysis of Methods for the selection of spectral models).
The results are shown in Fig.~\ref{Fig_spectrum} and
reported in Table~\ref{tab1}.
Two radiation components are identified by our spectral fittings,
including a thermal emission at around $200 \text{--} 500\,{\rm keV}$ and
a non-thermal emission spanning a wide energy range.
According to Fig.~\ref{Fig_spectrum},
the thermal component becomes dominant only in the time interval of $1.73 \text{--} 3.25\,{\rm s}$ and for $\mathord{\sim} 300\,{\rm keV}$ emission.
Importantly,
the thermal component dominates the $100 \text{--} 300\,{\rm keV}$ emission in $2.07 \text{--} 3.25\,{\rm s}$,
while the QPO modulation feature appears.
This indicates that the photosphere of GRB~240825A,
which produces the thermal emission component,
suffers from QPO modulation,
especially in the phase of $2.07 \text{--} 3.25\,{\rm s}$.

\begin{figure}
    \centering
\includegraphics[width=0.5\textwidth]{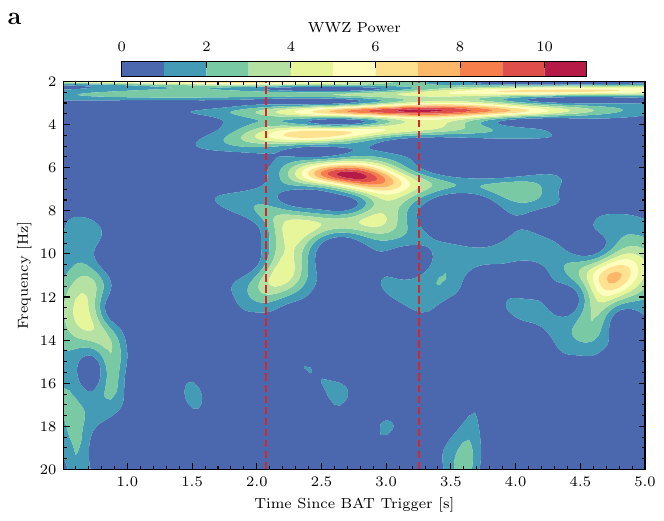}
\includegraphics[width=0.46\textwidth]{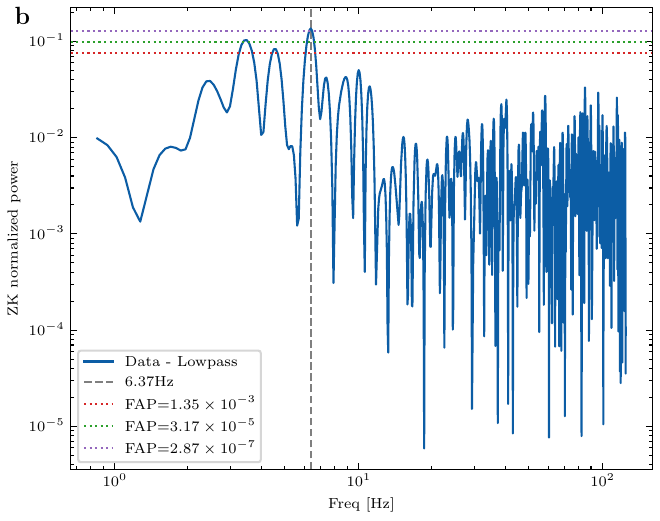}
\caption{Searches for the QPO signal in $100 \text{--} 300\,{\rm keV}$ light-curve.
(\textbf{a}) The WWZ power in the time-frequency space is shown.
A periodic signal at $\mathord{\sim} 6\,{\rm Hz}$ is clearly observed in the time interval of $2.07 \text{--} 3.25\,{\rm s}$,
which is consistent with the analysis in the  Fig.~\ref{Fig_multi_lc}\textbf{b}.
(\textbf{b}) The GLSP power (blue line)
for the detrended light curve,
i.e., the raw light curve
by subtracting its low-pass component,
during $2.07 \text{--} 3.25\,{\rm s}$ is shown.
A QPO signal peaking at $6.37\,{\rm Hz}$ is evident in this panel.
The FAPs with $1.35 \times 10^{-3}$, $3.17 \times 10^{-5}$, and $2.87 \times 10^{-7}$ are marked with red, green, and purple dashed lines in the right panel, respectively. 
It is shown that the $6.37\,\mathrm{Hz}$ QPO signal exhibits an FAP $< 2.87 \times 10^{-7}$,
which corresponds to a confidence level higher than $5\sigma$.}\label{Fig_wwz}
\end{figure}

We performed the detailed search for the QPO signals in the $100 \text{--} 300\,{\rm keV}$ light-curve
based on the weighted wavelet transform (WWZ) methodology\cite{1996AJ....112.1709F}.
The WWZ power in the time-frequency space is illustrated in  Fig.~\ref{Fig_wwz}\textbf{a}.
A strong QPO signal,
with an average period of $\mathord{\sim} 6\,{\rm Hz}$,
is obvious in the time slice of $\mathord{\sim} 2.07 \text{--} 3.25\,{\rm s}$.
We further perform the variability analysis by utilizing the generalized Lomb-Scargle periodogram (GLSP)\cite{2009A&A...496..577Z} on the $100 \text{--} 300\,{\rm keV}$ light-curve
during the time slice of $2.07 \text{--} 3.25\,{\rm s}$.
The result is shown in Fig.~\ref{Fig_wwz}\textbf{b},
where the blue solid line is the GLSP power of the 
detrended light curve (i.e., the raw light curve with its low-pass component subtracted) 
and the dashed lines with red, green, 
and purple depict the false alarm probabilities (FAPs) of $1.35 \times 10^{-3}$, $3.17 \times 10^{-5}$, and $2.87 \times 10^{-7}$, respectively.
The $6.37\,{\rm Hz}$ QPO exhibits a ${\rm FAP} < 2.87 \times 10^{-7}$, 
corresponding to a confidence level larger than $5\sigma$.
To indicate the QPO modulation on the $100 \text{--} 300\,{\rm keV}$ light-curve,
we plot a sinusoidal curve (with frequency of $6.33\,\mathrm{Hz}$) superimposed on a low-pass background light-curve (green dashed line)
in Fig.~\ref{Fig_multi_lc}\textbf{b} (see Methods).
There are seven full-circles within the phase of $2.07 \text{--} 3.25\,{\rm s}$.
GRB~240825A is the only one burst identified with a high-confidence level's low-frequency QPO signal in the thermal emission.

\begin{figure}
\centering
\includegraphics[width=0.8\textwidth]{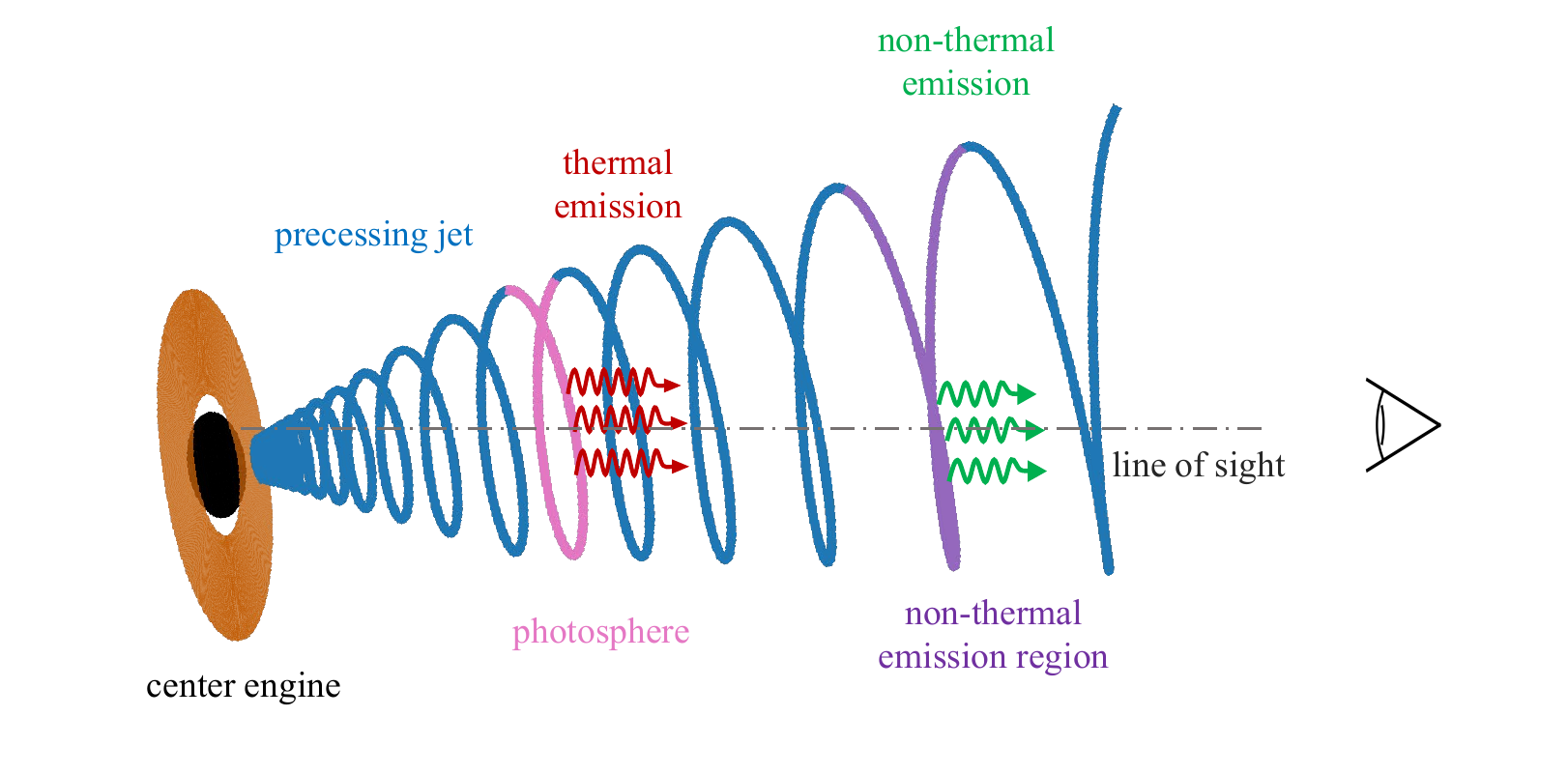}
\caption{Schematic illustration of a precessing jet
that produces periodically modulated emission.
A precessing jet is launched from the central engine.
The radius of the precessional circle for the jet flow increases as the jet expansion
and thus the circular period increases with time owing to the conservation of the angular momentum.
The thermal emission is from the photosphere of the jet
and the non-thermal emission is formed in the region beyond the photosphere.
Then, the QPO modulation on the thermal and non-thermal emission is exhibited with short and long periods, respectively.}
\label{Fig_model}
\end{figure}

\begin{figure}
\centering
\includegraphics[width=0.45\textwidth]{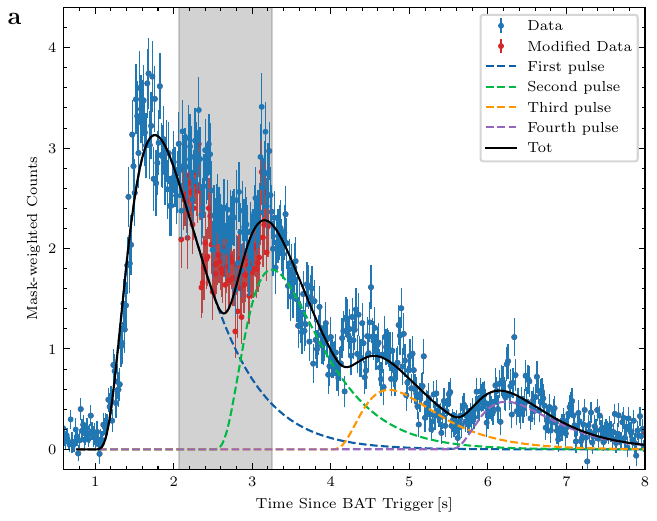}
\includegraphics[width=0.45\textwidth]{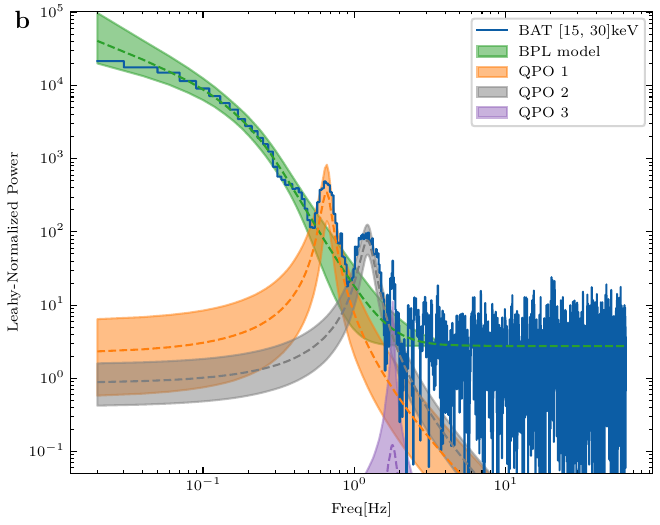}
\caption{Variability analysis on the non-thermal emission.
(\textbf{a}). It shows the $15 \text{--} 30\,\mathrm{keV}$ light-curve (blue dots).
The grey shaded area marks the time interval
during which clear periodic modulation on the thermal emission is found
and the thermal emission
dominates $\sim 100\text{--}300\,\mathrm{keV}$ emission.
The black line is the result of our light-curve fitting
on the $15 \text{--} 30\,\mathrm{keV}$ emission with periodic FRED pulses (see Methods),
where the blue, green, orange, and purple dashed-lines are the first, second, third, and fourth pulse, respectively.
The data during $2.0 \text{--} 3.0\,\mathrm{s}$ is not involved in our light-curve fitting
since the thermal component makes a significant contribution to the $15 \text{--} 30\,\mathrm{keV}$ emission in this phase.
However, the modified data (red dots), of which the thermal component is subtracted, 
are well consistent with the result of the light-curve fitting. 
(\textbf{b}). The Fourier periodogram analysis on the $15\text{--}30\,\mathrm{keV}$ emission during [-10,40]$\,\mathrm{s}$ (see Extended Data Table~\ref{exttab3} of Methods for the detail analysis).
The $\sim 0.63\,\mathrm{Hz}$ QPO (orange line) and its associated harmonics $0.63 \times 2\,\mathrm{Hz}$ QPO (gray line) and $0.63 \times 3\,\mathrm{Hz}$ QPO (purple line) are evident in this panel.
The shaded region indicates the $1\sigma$ confidence interval for the corresponding component.}
\label{Fig_LSP_syn}
\end{figure}

In short, our spectral and variability analyses reveal
a strong low-frequency QPO modulation
on the photosphere of GRB~240825A.
Low-frequency QPOs, on the order of $\mathord{\sim} 10 \text{--} 100\,\mathrm{Hz}$,
are predicted in several models.
For example, the precession of the jet can cause the change of the jet direction and thus QPO modulation on the light-curve \cite{2006A&A...454...11R,2010A&A...516A..16L,2024Univ...10..438H}.
If the precession is responsible for the QPO modulation on the photosphere,
one can expect a lower frequency QPO modulation on the non-thermal emission.
The reason is as follows (see the schematic illustration in Fig.~\ref{Fig_model}):
(1) the non-thermal emission region is beyond the photosphere
(2) the angular momentum of jet flow remains the same for its different precessional circles.
We also perform the light-curve
fitting on the $15\text{--}30\,{\rm keV}$ emission with periodic FRED pulses (see Methods),
of which the pulse's profile is the same but with different amplitude and beginning time.
The result is shown in Fig.~\ref{Fig_LSP_syn}\textbf{a} with a black line,
where the green, orange, red, and purple dashed-lines are
the first, second, third, and fourth pulse, respectively.
The periodic FRED pulses well depict the light-curve of the non-thermal emission with a period of $1.49^{+0.01}_{-0.01}\,\mathrm{s}$.
Fig.~\ref{Fig_LSP_syn}\textbf{b} presents the Fourier periodogram analysis on the $15\text{--}30\,\mathrm{keV}$ emission during $[-10, 40]\,\mathrm{s}$.
The $\mathord{\sim} 0.63\,\mathrm{Hz}$ QPO and its associated harmonics $0.63 \times 2\,\mathrm{Hz}$ and $0.63 \times 3\,\mathrm{Hz}$ QPOs is evident (see Extended Data Table~\ref{exttab3} of Methods for the periodogram analysis).
The results from the variability analysis and the light-curve fitting indicate a QPO in the non-thermal emission.

The facts of $6.37\,\mathrm{Hz}$ QPO in the thermal emission and $0.67\,\mathrm{Hz}$ QPO in the non-thermal emission are together to demonstrate a precessing jet in GRB~240825A.

\clearpage


\begin{table}
\renewcommand\arraystretch{1.5}
\centering
\caption{Results of the spectral analysis on GRB~240825A in four time slices.}
\label{tab1}
\begin{tabular}{|c|cccc|}
\hline
\hline
 & $[1.00, 1.73]\,{\rm s}$ & $[1.73, 2.07]\,{\rm s}$ & $[2.07, 3.25]\,{\rm s}$ & $[3.25, 8.00]\,{\rm s}$ \\
\hline
\hline
k$T_{\rm c}$ (keV) & $30.68^{+7.50}_{-4.58}$ & $25.32^{+2.32}_{-1.83}$ & $30.19^{+0.72}_{-0.66}$ & $16.05^{+2.05}_{-7.33}$ \\
$\alpha1$ & $-0.60^{+0.05}_{-0.07}$ & $-0.68^{+0.05}_{-0.05}$ & $-0.81^{+0.08}_{-0.08}$ & $-0.85^{+0.06}_{-0.08}$\\
$\alpha2$ & $-1.57^{+0.17}_{-0.20}$ & $-1.92^{+0.06}_{-0.05}$ & $-1.70^{+0.02}_{-0.02}$ & $-1.77^{+0.03}_{-0.03}$\\
$\beta$ & $-2.88^{+0.19}_{-0.30}$ & $-2.85^{+0.28}_{-0.48}$ & $-3.05^{+0.16}_{-0.17}$ & $-2.99^{+0.21}_{-0.25}$\\
$\log_{10}{(E_{\rm b}/\mathrm{keV})}$ & $2.09^{+0.06}_{-0.08}$ & $1.93^{+0.04}_{-0.05}$ & $1.57^{+0.05}_{-0.05}$ & $1.39^{+0.06}_{-0.06}$\\
$\log_{10}{(E_{\rm p}/\mathrm{keV})}$ & $3.17^{+0.14}_{-0.10}$ & $4.63^{+0.28}_{-0.27}$ & $4.58^{+0.07}_{-0.08}$ & $4.52^{+0.12}_{-0.14}$ \\
$F_{\rm CPL\text{--}mBB}$ ($\rm erg/cm^{2}/s$) & $2.26^{+0.39}_{-0.45} \times 10^{-5}$ & $2.06^{+0.23}_{-0.23} \times 10^{-5}$ & $2.76^{+0.07}_{-0.07} \times 10^{-5}$ & $7.33^{+1.13}_{-1.17} \times 10^{-7}$ \\
$F_{\rm 2SBPL}$ ($\rm erg/cm^{2}/s$) & $4.37^{+0.45}_{-0.42} \times 10^{-5}$ & $4.22^{+0.33}_{-0.30} \times 10^{-5}$ & $3.11^{+0.13}_{-0.13} \times 10^{-5}$ & $5.99^{+0.37}_{-0.39} \times 10^{-6}$ \\
PG-stat/dof & $351.96/325$ & $373.57/355$ & $382.69/355$ & $371.22/355$\\

\hline
\hline
\end{tabular}
\begin{tablenotes}
\raggedleft
\item The energy fluxes are calculated between $1\,{\rm keV}$ and $10\,{\rm MeV}$.
    All errors represent the $1 \sigma$ uncertainties.
\end{tablenotes}
\end{table}

\clearpage
\bmhead{Acknowledgements}
We acknowledge the use of the Fermi and Swift archive's public data.
We thank all the contributors of the open source software/code we use.
We thank Bing Zhang for his helpful discussion on this work.
This work is supported by the National Natural Science Foundation of China (grant Nos. 12273005, 12133003,
and U1938201),
the Guangxi Science Foundation (grant Nos. 2018GXNSFFA281010 and 2018GXNSFGA281007),
and China Manned Spaced Project (CMS-CSST-2021-B11).

\section*{Declarations}
\begin{itemize}
\item Funding:
This work is supported by the National Natural Science Foundation of China (grant Nos. 12273005, 12494575, and 12133003), the National Key R\&D Program of China (grant No. 2023YFE0117200 and 2024YFA1611700), the special funding for Guangxi Bagui Youth Scholars, and the Guangxi Talent Program (``Highland of Innovation Talents'').

\item Conflict of interest:
The authors declare that they have no conflict of interest.

\item Data availability:
Fermi-GBM/Fermi-LAT data are publicly available from the Fermi Public Data Archive (\url{https://heasarc.gsfc.nasa.gov/FTP/fermi/data/}).
Swift-BAT data are publicly available from the Swift archive website (\url{https://www.swift.ac.uk/archive/ql.php}).

\item Code availability:
All software used for the analyses is publicly available.
HEASoft which include Xspec (\url{https://heasarc.gsfc.nasa.gov/lheasoft/}),
GBM-Data-Tools (\url{https://fermi.gsfc.nasa.gov/ssc/data/analysis/gbm/}),
Fermitools which include GTBURST (\url{https://github.com/fermi-lat/Fermitools-conda}),
BXA (\url{https://github.com/JohannesBuchner/BXA/}),
UltraNest (\url{https://github.com/JohannesBuchner/UltraNest}),
Python package for WWZ (\url{https://github.com/skiehl/wwz}),
PyAstronomy (\url{https://github.com/sczesla/PyAstronomy}),
Stingray (\url{https://github.com/StingraySoftware/stingray}),
Emcee (\url{https://github.com/dfm/emcee}), 
Dynesty (\url{https://github.com/joshspeagle/dynesty}), 
Bilby (\url{https://github.com/bilby-dev/bilby}),
SciPy (\url{https://github.com/scipy/scipy}),
All plots are generated by Matplotlib (\url{https://github.com/matplotlib/matplotlib}).

\item Author contribution:
The study is leaded by Da-Bin Lin
and initiated by both Da-Bin Lin and Guo-Yu Li.
The contributions to write the manuscript are Guo-Yu Li,
Da-Bin Lin, Zhi-Lin Chen, Bao-Quan Huang, Tong Liu, and En-Wei Liang.
\end{itemize}

\newpage
\clearpage
\clearpage
\setcounter{figure}{0}
\setcounter{table}{0}
\captionsetup[table]{name={\bf Extended Data Table}}
\captionsetup[figure]{name={\bf Extended Data Fig.}}
\section*{Methods}\label{sec:methods}

\subsection*{Observations and Data Analysis}\label{subsec:obs}
On 25 August 2024 at 15:52:59.84 UTC (hereafter $T_{0}$),
the Swift Burst Alert Telescope (BAT) was triggered by GRB~240825A\cite{2024GCN.37274....1G} and then the location of this burst was reported.
Shortly thereafter, the Fermi Gamma-ray Burst Monitor (GBM) detected GRB~240825A at $\mathord{\sim} T_{0} + 0.25\,{\rm s}$.
In addition,
GRB~240825A was also detected by several high-energy missions including Fermi-LAT\cite{2024GCN.37288....1D},
Konus-Wind\cite{2024GCN.37302....1F},
AstroSat/CZTI\cite{2024GCN.37298....1J},
and INTEGRAL/SPI\cite{2024GCN.37537....1P}.
The host galaxy of GRB~240825A was reported with a redshift of $z = 0.659$\cite{2024GCN.37293....1M}.

\textbf{Fermi data analyses.\;}
The Fermi data of GRB~240825A were obtained from the Fermi Public Data Archive (\url{https://heasarc.gsfc.nasa.gov/FTP/fermi/data/}).
We select two sodium iodide (NaI) detectors (n6 $\&$ n7) and one bismuth germanium oxide (BGO) detector (b1) with the smallest viewing angles with respect to the GRB source direction.
The Fermi-GBM light-curve was obtained based on the time-tagged events (TTE) data.
The light-curves in different energy ranges are extracted from the TTE data by the GBM-Data-Tools\cite{GbmDataTools}.
To create the GBM light-curve in Fig.~\ref{Fig_multi_lc},
we first selected the photons in the b1 detector with an energy range of $1,000 \text{--} 40,000\,{\rm keV}$,
in which all the photons are binned in time of $0.016\,{\rm s}$ without removing the photons from the background.
Then,
we fitted the background with a polynomial of the one order of \emph{BackgroundFitter}.
The source light-curve was obtained by subtracting the fitted light-curve of the background.
Finally, the light-curve was normalized with its maximum count.

\textbf{Swift data analyses.\;}
We obtained the BAT observational data from the Swift archive website (\url{https://www.swift.ac.uk/archive/ql.php}).
Using the HEASoft tools\cite{2014ascl.soft08004N} (version 6.30.1),
we obtained the BAT light-curves following the standard analysis procedures (\url{https://www.swift.ac.uk/analysis/bat/}).
We checked the energy conversion, generated a quality map, defined hot pixels, and applied mask weighting.
Finally,
the mask-weighted light-curves are generated by \emph{batbinevt}.
The mask-weighted light-curve,
with the background removed by ray tracing techniques,
yields a clean light-curve.
The light-curves in Fig.~\ref{Fig_multi_lc} are normalized with their maximum mask-weighted count, respectively.

\subsection*{Spectral Analysis}\label{subsec:spectral}
\textbf{Spectral Models.\;}
Our spectral analysis involves four spectral models,
i.e., 2SBPL, CPL-mBB, Band function, and CPL.

2SBPL is a double smoothly broken power-law model (\cite{2018A&A...613A..16R}) defined as
\begin{multline}
\begin{split}
\label{eq:2sbpl}
    N(E) &= A E_{\rm b}^{\alpha_{1}} \bigg[\bigg[\bigg(\frac{E}{E_{\rm b}}\bigg)^{-\alpha_{1}n_{1}}+\bigg(\frac{E}{E_{\rm b}}\bigg)^{-\alpha_{2}n_{1}}\bigg]^{\frac{n_{2}}{n_{1}}}\\
    &+ \bigg(\frac{E}{E_{\rm j}}\bigg)^{-\beta n_{2}}\bigg[\bigg(\frac{E_{\rm j}}{E_{\rm b}}\bigg)^{-\alpha_{1} n_{1}} + \bigg(\frac{E_{\rm j}}{E_{\rm b}}\bigg)^{-\alpha_{2} n_{1}}\bigg]^{\frac{n_{2}}{n_{1}}}\bigg]^{-\frac{1}{n_{2}}}
\end{split}
\end{multline}
where ${E_{\rm j}} = {E_{\rm p}}{[ - ({\alpha _{2}} + 2)/(\beta  + 2)]^{1/[(\beta  - {\alpha_{2}}){n_{2}}]}}$ and the free parameters are the amplitude $A$, 
the break energy $E_{\rm b}$, 
the peak energy $E_{\rm p}$, 
the photon index $\alpha_1$ below $E_{\rm b}$, 
the photon index $\alpha_2$ between $E_{\rm b}$ and $E_{\rm p}$, and the photon index $\beta$ above $E_{\rm p}$.
The smoothness parameters $n_{1} = 5.38$ and $n_{2} = 2.69$.
The 2SBPL is used to describe the non-thermal component of the spectrum.

CPL-mBB spectral model is used to describe the spectrum of a non-dissipative photospheric emission \cite{2022ApJ...934L..22D}.
It is composed of a multicolor blackbody with a cut-off power-law distribution of temperature,
i.e.,
\begin{equation}
\label{eq:mbb}
	N(E)=\int_{T_{\rm{min}}}^{T_{\rm max }}  \frac{dB(T)}{dT} \frac{E^2}{\exp(E/kT)-1}dT,
\end{equation}
where $B(T)=f(T)/(T^4 \pi^4/15)$ and $f(T)=f_{\rm max} (T/T_{\rm c})^q \exp[-(T/T_{\rm{c}})^s]$.
The distribution function of temperature $f(T)$ with $T_{\rm c} \gg T_{\rm max }$
is reduced to $f(T)=f_{\rm max} (T/T_{\rm c})^q$.
The CPL-mBB with $q=2.8$ and $s=0.9$ well describes the spectral morphology of the photospheric emission,
and is used in this paper.

Band function is widely used to fit and describe the radiation spectrum of GRBs\cite{1993ApJ...413..281B}.
Its functional form is described as follows,
\begin{equation}
\label{eq:band}
	N(E) = K \left\{
	\begin{array}{ll}
		\left( \frac{E}{100 \ \rm keV} \right)^{\alpha} \exp \left( -\frac{E}{E_{\rm{c}}}  \right), & E < (\alpha - \beta) E_{\rm{c}},   \\
		\left( \frac{E}{100 \ \rm keV} \right)^{\beta} \exp\left(\beta - \alpha\right)\left[ \frac{(\alpha - \beta) E_{\rm{c}}}{100 \ \rm keV}
		\right]^{\alpha - \beta}, & E \geq (\alpha - \beta) E_{\rm{c}},
	\end{array}
	\right.
\end{equation}
where $K$ is the normalization constant, and $\alpha$, $\beta$, and $E_{\rm c}$ are the low-energy photon spectral index,
high-energy photon spectral index,
and the break photon energy,
respectively.

The CPL model is defined as
\begin{equation}
\label{eq:cpl}
    N(E) = K E^{-\alpha} \exp \left(-\frac{E}{E_{\rm c}}  \right),
\end{equation}
where $K$ is the normalization constant, 
$\alpha$ is the power-law index, 
and $E_{\rm c}$ is the cutoff energy.

\textbf{Spectral analysis and the thermal component identification.\;}
The Python source package GTBURST\cite{2016zndo.....59783V} is used to obtain the GBM and LAT spectral files for GRB~240825A
based on the standard analysis process.
For GBM data, we selected background intervals before and after the burst to model the background with a polynomial fit.
For events of LAT,
we select the Transient type with a maximum zenith angle of $100^{\circ}$ and within $12^{\circ}$ along the orientation of GRB~240825A.
In our spectral fittings,
we use the spectral analysis software BXA\cite{2014A&A...564A.125B},
which connects the nested sampling algorithm UltraNest\cite{2021JOSS....6.3001B} with PyXspec\cite{2021ascl.soft01014G}.
In addition, the PG-statistic is adopted in these spectral fittings.
Data in the $30 \text{--} 40\,\mathrm{keV}$ energy range are excluded to avoid due to the iodine K-edge.

We performed spectral analyses on four time intervals: 
$1.0 \text{--} 1.73\,\mathrm{s}$ (BIN1), 
$1.73 \text{--} 2.07\,\mathrm{s}$ (BIN2), 
$2.07 \text{--} 3.25\,\mathrm{s}$ (BIN3), 
and $3.25 \text{--} 8.0\,\mathrm{s}$ (BIN4).
It is found that the Band function and 2SBPL model could not well fit the observations for these time intervals, especially for BIN3-4. 
Similar results are also reported in ref.~\cite{2025ApJ...984L..45Z,2025ApJ...985L..30W}.
Then, we test different combinations of spectral models,
involving 2SBPL$+$CPL-mBB and Band$+$CPL models.
The 2SBPL$+$CPL-mBB model emerges as an optimal one for the following two reasons.

Firstly, we perform the spectral fittings for these time intervals with models of Band function,
2SBPL, 2SBPL$+$CPL-mBB, and Band$+$CPL.
The fitting results are reported in Extended Data Table~\ref{exttab1}.
Given model complexity and different numbers of free parameters, 
we employed the Akaike Information Criterion (AIC; \cite{1974ITAC...19..716A}) and Bayesian Information Criterion
(BIC; \cite{schwarz1978estimating})
to compare the goodness-of-fit across different models.
These metrics quantify the trade-off
between model fit quality and parameter efficiency, 
where lower values indicate better-balance models.
We calculate the values of
$\rm \Delta{AIC}=AIC_{c}-AIC_{2SBPL+CPL-mBB}$ and
$\rm \Delta{BIC}=BIC_{c}-BIC_{2SBPL+CPL-mBB}$
for other candidate models relative to the ``2SBPL$+$CPL-mBB" benchmark model.
According to simple rules of thumb, 
the situations of $-2\leqslant \Delta{\rm AIC}\leqslant 2$ or $-2\leqslant \Delta{\rm BIC}\leqslant 2$ indicate comparable support for the candidate and benchmark models,
the situations with $\Delta{\rm AIC}>10$ or $\Delta{\rm BIC}>10$ has essentially no support for the candidate model,
and the situations with $\Delta{\rm AIC}<-10$ or $\Delta{\rm BIC}<-10$ has essentially no support for the benchmark model\cite{doi:10.1177/0049124104268644,2015EPJC...75....5S}.
The Extended Data Table~\ref{exttab1} reports that the values of $\Delta{\rm AIC}$ and $\Delta{\rm BIC}$ are all significantly larger than
10 for both BIN2-4,
which strongly supports the spectral model of 2SBPL$+$CPL-mBB.
For the time interval of BIN1,
the values of $\Delta{\rm AIC}$ and $\Delta{\rm BIC}$
tend to support the spectral model of 2SBPL rather than other models.
This may be related to that the contribution of the thermal component in this time interval
is negligible,
which is consistent with the spectral fitting result
based on the 2SBPL$+$CPL-mBB model for this time interval.

Secondly and most importantly,
the distinct variabilities between $15 \text{--} 30\,\mathrm{keV}$ and $100 \text{--} 300\,\mathrm{keV}$ (see Extended Data Fig.~\ref{Fig_wwz_compare}) indicate different origin for these two bands.
That is to say, the emission in the $15 \text{--} 30\,\mathrm{keV}$ and $100 \text{--} 300\,\mathrm{keV}$ bands should be dominated by different spectral models.
For the Band$+$CPL and 2SBPL$+$CPL-mBB models,
we calculate the contribution ratio of the spectral components for both $15 \text{--} 30\,\mathrm{keV}$ and $100 \text{--} 300\,\mathrm{keV}$ bands.
The results are reported in
Extended Data Table~\ref{exttab1},
i.e.,
$\log_{10}({F_{\rm 15-30,Band}}/{F_{\rm 15-30,CPL}})$ and $\log_{10}({F_{\rm 100-300,Band}}/{F_{\rm 100-300,CPL}})$ vs. $\log_{10}({F_{\rm 15-30,2SBPL}}/{F_{\rm 15-30,CPL-mBB}})$ and $\log_{10}({F_{\rm 100-300,2SBPL}}/{F_{\rm 100-300,CPL-mBB}})$.
One can find that the Band spectral component makes
the main contribution to both
$15 \text{--} 30\,\mathrm{keV}$ and $100 \text{--} 300\,\mathrm{keV}$ bands
in the spectral fittings based on Band$+$CPL model.
This is inconsistent with the results from the variability analysis.
In the spectral fittings with 2SBPL$+$CPL-mBB model,
the main contribution to the $15 \text{--} 30\,\mathrm{keV}$ emission are the 2SBPL spectral component for these four time intervals.
However, the contribution to the $100 \text{--} 300\,\mathrm{keV}$ emission is different.
For the time intervals of BIN1-2,
the CPL-mBB and 2SBPL spectral components make comparable contributions,
which is consistent with the spectral model comparison based on AIC or BIC criteria.
For the time intervals of BIN3,
the CPL-mBB spectral component makes the main contribution.
For the time intervals of BIN4,
the 2SBPL spectral component makes the main contribution.
The dominance of the CPL-mBB spectral component at the $100 \text{--} 300\,\mathrm{keV}$ emission during BIN3 is consistent with the fact that
the $6.37\,\mathrm{Hz}$ QPO is only detected in this interval
with high significance. 
This provides direct evidence that the reported $6.37\,\mathrm{Hz}$ QPO is associated with the thermal emission.

In summary, both AIC and BIC criteria and the variability analysis favor the spectral model of the 2SBPL$+$CPL-mBB model for the time intervals of BIN2-4, especially BIN3.

\subsection*{WWZ Analysis\;}\label{subsec:wwz}
\begin{figure}
\centering
\includegraphics[width=1.0\textwidth]{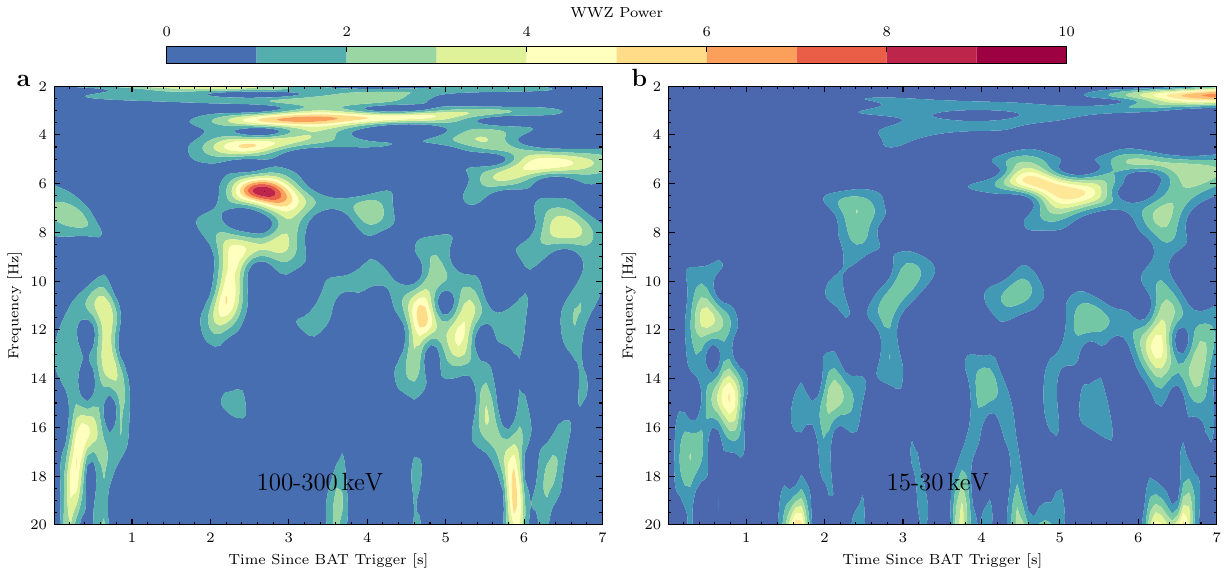}
\caption{Comparison of WWZ time-frequency analyses for light curves in different energy bands over the same time and frequency intervals: (\textbf{a}) $100 \text{--} 300\,\mathrm{keV}$, (\textbf{b}) $15 \text{--} 30\,\mathrm{keV}$.}
    \label{Fig_wwz_compare}
\end{figure}

Periodic modulation may occur in some phases,
and the weighted wavelet Z-transform (WWZ) method is well-suitable for detecting the power of any dominant periodic modulation and its corresponding time span in the light curve\cite{1996AJ....112.1709F}.
In this method, a WWZ map is created by convolving the light curve with a time and frequency-dependent kernel, and then decomposing the data into time and frequency domains.
The WWZ map for a light-curve $y(t)$ is obtained with
\begin{equation}
\label{eq:wwz_W}
	W(\omega, \tau) = \omega^{1/2} \int y(t) f^{\ast}(\omega,\tau;t) d{t}
\end{equation}
by varying the value of the time-shift $\tau$ and the angular frequency $\omega$. 
Here, $f^{\ast}$ is the complex conjugate of the Morlet wavelet kernel $f$ as 
\begin{equation}
\label{eq:wwz_f}
	f(\omega,\tau;t) = \mathrm{e}^{[i \omega(t - \tau) - c \omega^{2}(t - \tau)^{2}]},
\end{equation}
$\omega$ is the angular frequency for the searched quasi-periodic signals,
and $c$ is a decay constant controlling wavelet width.
To perform the WWZ analysis on GRB~240825A, 
we employ a publicly available Python WWZ analysis package\cite{2023ascl.soft10003K}. 
We search for periodic signals with frequencies in the $2\text{--}20\,\mathrm{Hz}$ range and set $c = 1 / (32 \pi^{2})$ to ensure the decay window covers two cycles.
The results from the analysis of $100\text{--}300\,\mathrm{keV}$ emission are shown in Fig.~\ref{Fig_wwz}\textbf{a} and Extended Data Fig.~\ref{Fig_wwz_compare}\textbf{a},
where a distinct periodic signal with a frequency of $\mathord{\sim} 6\,{\rm Hz}$ during 2.07 to 3.25$\,{\rm s}$ can be found.
In Extended Data Fig.~\ref{Fig_wwz_compare}\textbf{b}, the results from the analysis of $15 \text{--}30\,\mathrm{keV}$ emission is shown.
Comparing Extended Data Fig.~\ref{Fig_wwz_compare}\textbf{a} and Extended Data Fig.~\ref{Fig_wwz_compare}\textbf{b},
one can find that the variabilities of $15\text{--}30\,\mathrm{keV}$ 
$100 \text{--} 300\,\mathrm{keV}$ emission are very different
and the $\mathord{\sim} 6\,{\rm Hz}$ QPO only appears in the $100\text{--}300\,\mathrm{keV}$ emission.

\begin{figure}
	\centering
	\includegraphics[width=0.48\textwidth]{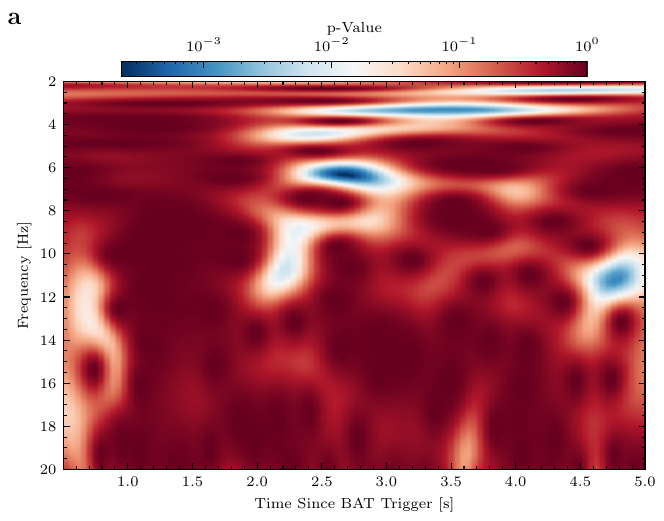}
	\includegraphics[width=0.48\textwidth]{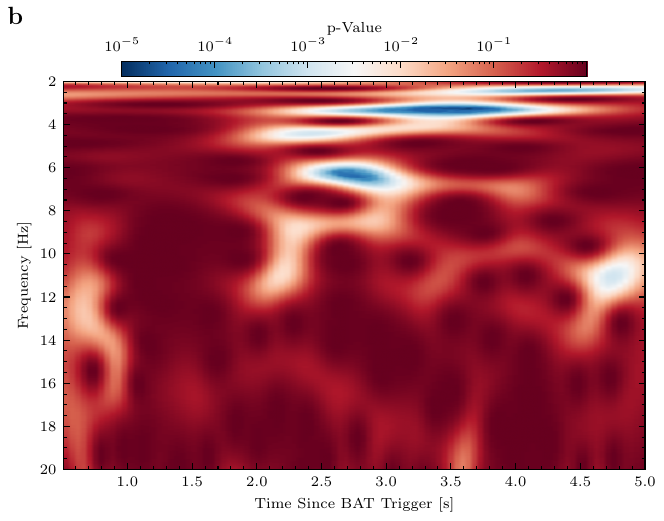}
	\caption{The p-value distributions of the WWZ, derived from $10^{5}$ Gaussian white noise simulations, 
    are shown for time bin sizes of $8\,{\rm ms}$ (panel (\textbf{a})) and $4\,{\rm ms}$ (panel (\textbf{b})).
	}
	\label{Fig_wwz_pvalue}
\end{figure}

To quantify the significance of detected periodic signals, 
we perform Monte Carlo simulations by generating $10^{5}$ Gaussian white noise light curves
and calculate the corresponding WWZ power $W_{\text{sim}}$.
For each frequency-time bin, we calculate the $p$-value,
which is the fraction of noise realizations exhibiting WWZ power exceeding the observed value $W_{\text{obs}}$ (see Fig.~\ref{Fig_wwz}\textbf{a}), i.e.,
\begin{equation}
\label{eq:wwz_p}
	p = \Pr\left[ W_{\text{sim}}(\omega, \tau) > W_{\text{obs}}(\omega, \tau) \right].
\end{equation}
Here, $\Pr$ is the probability function for the situation of $W_{\text{sim}}(\omega, \tau) > W_{\text{obs}}(\omega, \tau)$.
The results from the analysis of $4\,\mathrm{ms}$ and $8\,\mathrm{ms}$ binned data are showed in the Extended Data Fig.~\ref{Fig_wwz_pvalue}.
One should note that the $p$-value gives us an estimate of the probability that the simulations, 
which are based on the null-hypothesis $H_0$ (i.e., the Gaussian noise), 
result in a signal with WWZ power equal to or stronger than that of the real data. 
Thus, it tests the null hypothesis ($H_{\rm 0}$) that the observed signal originates from Gaussian noise against the alternative hypothesis ($H_{1}$) of the quasi-periodic signal.
The Extended Data Fig.~\ref{Fig_wwz_pvalue} indicates
rejection of white noise ($H_{\rm 0}$ ) origin for periodic signals of $\mathord{\sim} 6\,{\mathrm{Hz}}$ during the time interval of $\mathord{\sim} 2.07 \text{--} 3.25\,\mathrm{s}$ with $p < 0.00135$
or $>3\sigma$ significance level.

\subsection*{Generalized LSP Analysis\;}\label{subsec:GLSP}
The Lomb-Scargle periodogram (LSP) is a widely used tool in period searches and frequency analysis of time series\cite{1976Ap&SS..39..447L,1982ApJ...263..835S}. 
Its generalized version (GLSP\cite{2009A&A...496..577Z}) extends this framework by incorporating a floating mean and weighted least-squares fitting\cite{2018ApJS..236...16V}. 
The GLSP models the data as
\begin{equation}
\label{eq:lsp_y}
	y(t) = a \cos{\omega t} + b \sin{\omega t} + c, 
\end{equation}
and then minimizes the weighted residual sum of squares
\begin{equation}
\label{eq:lsp_chi2}
	\chi^{2} = \sum_{i=1}^{N} w_{i}[y_{i} - y(t_{i})]^{2} ,
\end{equation}
where $y_{i}$ is the measurement at time $t_{i}$, $w_{i}=\frac{1}{\sigma_{i}} \sum_{i=1}^{N} \sigma_{i}^{2}$ is normalized weight, 
and $\sigma_{i}^2$ is the variance of the data $y_{i}$.
The GLSP power $P(\omega)$ at each angular frequency $\omega$ is defined as:
\begin{equation}
\label{eq:lsp_P}
	P(\omega) = \frac{\chi_{0}^{2} - \chi^{2}(\omega)}{\chi_{0}^{2}},
\end{equation}
where $\chi_{0}^{2}$ represents the residuals by fitting the data with a constant model (i.e., $a=0$ and $b=0$),
and serves as a baseline to quantify the proportional reduction in $\chi^{2}$ after introducing a sinusoidal component.
For Gaussian white noise and a given GLSP power $P_{\mathrm{n}}$,
the probability to obtain a GLSP power greater than $P_{\mathrm{n}}$ is 
\begin{equation}
\label{eq:lsp_Pr}
    \mathrm{Pr}(P > P_{\mathrm{n}}) = (1 - P_{\mathrm{n}})^{(N-3)/2},
\end{equation}
where $N$ is the number of observed data points.
The corresponding false-alarm probability (FAP) becomes:
\begin{equation}
\label{eq:lsp_FAP}
    \mathrm{FAP} = 1 - [1 - \mathrm{Pr}(P > P_{\mathrm{n}})]^{M},
\end{equation}
where $M = T \cdot \Delta{f}$,
$\Delta{f}$ is the bandwidth of the focused frequencies, 
and $T$ is the total duration of the analyzed observational data. 
The FAP represents the probability that at least one of the $M$ independent power values in a prescribed search band of a white-noise power spectrum exceeds $P_{\mathrm{n}}$.
For detailed calculation methods, 
please refer to ref.~\cite{2009A&A...496..577Z}.

\begin{figure}
	\centering
	\includegraphics[width=0.48\textwidth]{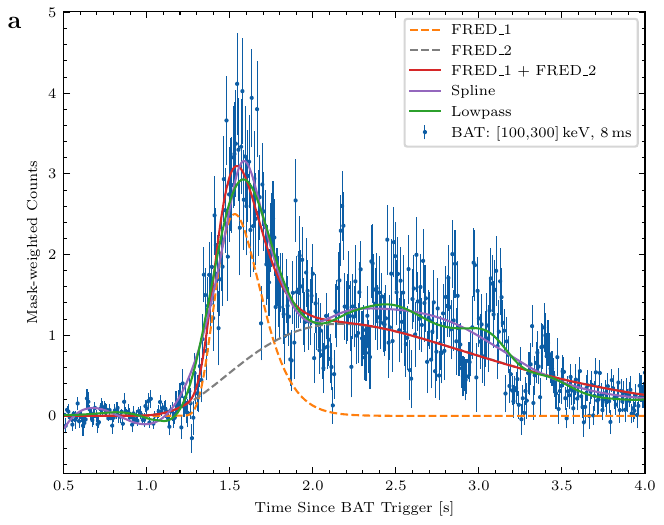}
	\includegraphics[width=0.48\textwidth]{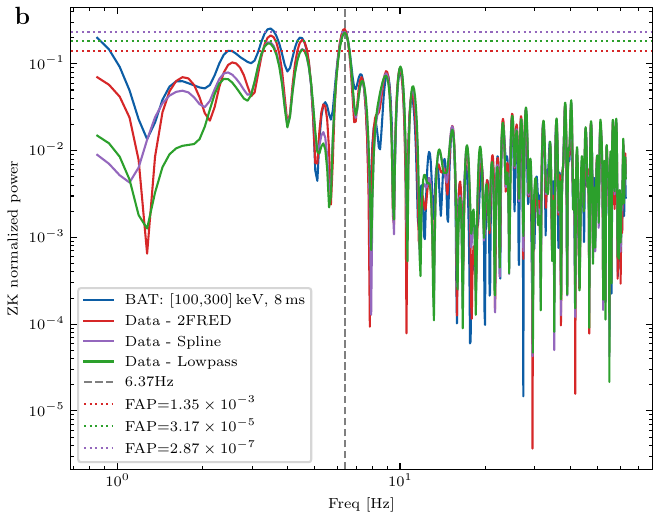}
	\includegraphics[width=0.48\textwidth]{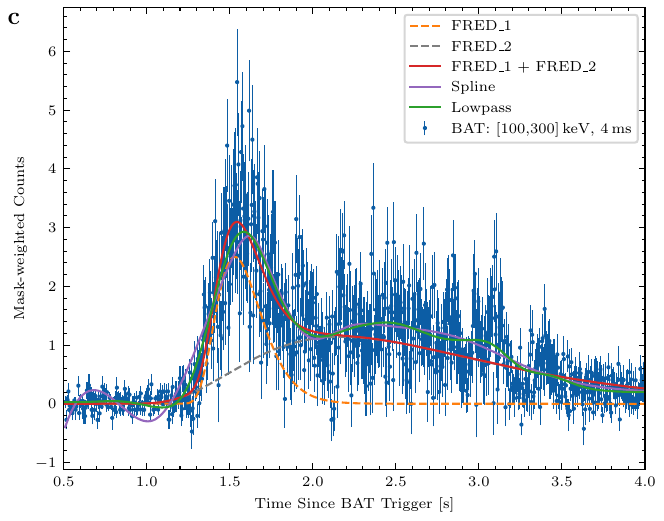}
	\includegraphics[width=0.48\textwidth]{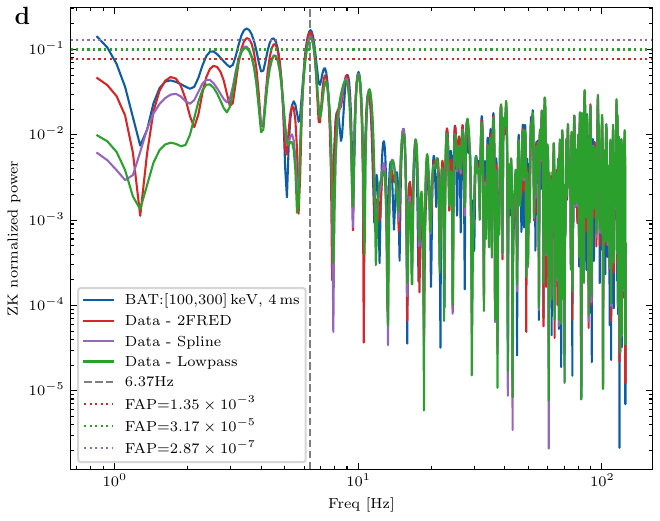}
	\caption{Panel (\textbf{a}) and (\textbf{c}): 
    Trend modeling of the $100 \text{--} 300\,\mathrm{keV}$ light curves. 
    Red, purple, and green solid lines represent the results of two independent FRED fits, 
    spline interpolation, 
    and low-pass filtering, 
    respectively.
    Panel (\textbf{b}) and (\textbf{d}): 
    Blue, red, purple, and green solid lines correspond to the GLSP of the original light curve, 
    GLSP after detrending with two FRED components, 
    GLSP after spline detrending, 
    and GLSP after detrending its low-pass component. 
    Red, green, and purple dashed lines denote FAP levels of $1.35 \times 10^{-3}$, 
    $3.17 \times 10^{-5}$, 
    and $2.87 \times 10^{-7}$, 
    respectively.
    Analysis based on $8\,\mathrm{ms}$ binned light curves (\textbf{a}, \textbf{b}); 
    analysis based on $4\,\mathrm{ms}$ binned light curves (\textbf{c}, \textbf{d}).
	}
	\label{Fig_LSP_detrend}
\end{figure}

Our WWZ analysis identifies a quasi-periodic modulation during $2.07\text{--}3.25\,\mathrm{s}$.
However, the trend of the $100 \text{--} 300\,\mathrm{keV}$ emission during this time interval and its effect on the estimation of the confidence significance is not removed.
In the following, we perform the GLSP analysis on the $100 \text{--} 300\,\mathrm{keV}$ emission during $2.07 \text{--} 3.25\,\mathrm{s}$
based on the Python source package PyAstronomy\cite{pya}.
The following three models on the light-curve trend are employed in our light-curve detrend of the $100\text{--}300\,\mathrm{keV}$ emission during $2.07\text{--}3.25\,\mathrm{s}$:
\begin{enumerate}
\item  Low-pass component, i.e., the light-curve filtered with a low-pass filter.
Here, we apply a Butterworth low-pass filter from SciPy with order $3$ and a cutoff frequency of $2\,\mathrm{Hz}$ on the $100 \text{--} 300\,\mathrm{keV}$ light-curve.
The resulting low-pass component is shown with a green dashed line in Fig.~\ref{Fig_multi_lc}\textbf{b}
and a green solid line in Extended Data Fig.~\ref{Fig_LSP_detrend}\textbf{a,c}.
\item FRED-pulse Modelling: 
The light-curve of $100\text{--}300\,\mathrm{keV}$ can be decomposed
into different pulses with fast-rising exponential decay (FRED) profile, e.g., Equation~\eqref{eq:FRED}.
For the $100\text{--}300\,\mathrm{keV}$ emission during $0\text{--}8\,\mathrm{s}$,
the light-curve is modelled with two FRED-pulses.
The best fitting results are
$\log_{10}(A)=5.41$,  
$\Delta=1.07$, 
$\tau=0.46$, 
$\zeta=5.77$ for the first FRED-pulse (orange dashed-line)
and $\log_{10}(A)=2.09$, 
$\Delta=0.61$, 
$\tau=1.56$, 
$\zeta=2.34$ for the second FRED-pulse
(gray dashed-line), 
and the sum of these two pulses is showed with the red solid line in Extended Data Fig.~\ref{Fig_LSP_detrend}\textbf{a,c}.
\item Smoothing the light-curve based on the spline method.
Here, the $UnivariateSpline$ in SciPy\cite{2020SciPy-NMeth} are employed with smoothing factors of $280$ for $4\,\mathrm{ms}$ binned light-curve and $70$ for $8\,\mathrm{ms}$ binned light-curve.
The resutls are showed with purple lines in Extended Data Fig.~\ref{Fig_LSP_detrend}\textbf{a,c}. 
\end{enumerate}

Based on the above three models on the light-curve trend, we detrend the light-curve of $100\text{--}300\,\mathrm{keV}$ emission during $2.07\text{--}3.25\,\mathrm{s}$
and estimate the GLSP power for the residual light-curve. 
The results are showed in the Extended Data Fig.~\ref{Fig_LSP_detrend}\textbf{b,d} with the same color as the corresponding trend in the left panels.
The GLSP power for the raw light-curve is also estimated and shown with a blue line in the Extended Data Fig.~\ref{Fig_LSP_detrend}{b,d}.
A significant quasi-periodic signal at $6.37\,\mathrm{Hz}$ can be found in these four estimation methods.
For the light curves with $4\,\mathrm{ms}$ bin, 
the confidence level of $6.37\,\mathrm{Hz}$ QPO exceeds $5\sigma$ with $\mathrm{FAP} < 2.87 \times 10^{-7}$for these four estimation methods.
For the light curves with $8\,\mathrm{ms}$ bin, 
the FAP of $6.37\,\mathrm{Hz}$ QPO is $\approx 2.87 \times 10^{-7}$,
and the details are shown in Extended Data Table~\ref{exttab2}.

\subsection*{Fourier periodogram Analysis\;}\label{subsec:FP}
The Fourier periodogram is widely employed in astronomical data analysis as a fundamental tool for detecting quasi-periodic features\cite{2022hxga.book....7B}. 
We address the detection of periodicity in noisy time series within a Bayesian framework\cite{2010MNRAS.402..307V,2013ApJ...768...87H} and compute Fourier periodograms using the Stingray\cite{matteo_bachetti_2024_13974481,2019ApJ...881...39H,bachettiStingrayFastModern2024} Python library. 
The power of periodogram is Leahy-normalized, 
and thus the mean power of a pure Poisson noise data is $2$\cite{1983ApJ...266..160L}.
It is noteworthy that the mask-weighted light curves of BAT about GRBs provide background-subtracted flux for the source. 
The deconvolution technique inherently introduces Gaussian errors rather than Poisson noise.

\begin{figure}
\centering
\includegraphics[width=1.0\textwidth]{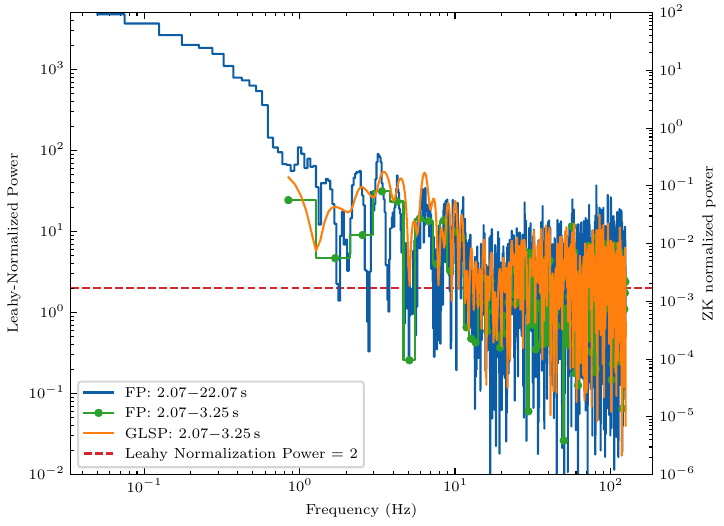}
\caption{Comparison of Leahy-normalized Fourier periodogram (FP) and ZK-normalized GLSP for $100\text{--}300\,\mathrm{keV}$ emission. 
Blue line: FP over $2.07\text{--}22.07\,\mathrm{s}$. 
Green line/datapoints: FP over $2.07 \text{--} 3.25\,\mathrm{s}$. 
Orange line: GLSP over $2.07\text{--}3.25\,\mathrm{s}$.}
    \label{Fig_FP_GLSP_compare}
\end{figure}

\textbf{Selection of Time Segments.\;}
One should note that both the frequency resolution $\Delta{f}$ and the minimum frequency $f_{\rm min}$ of a Fourier periodogram is associated with the duration $T$ of the focused time series,
i.e., $\Delta{f} = 1/T$ and $f_{\rm min} = 1/T$.
For example, the periodogram analysis of the $100 \text{--} 300\,\mathrm{keV}$ emission during $2.07\text{--}3.25\,\mathrm{s}$ has a frequency resolution of $\Delta{f} \sim 0.85\,\mathrm{Hz}$,
and thus the profile of the low-frequency QPOs ($\sim f_{\rm min}$, e.g., $6.37\,\mathrm{Hz}$ QPO) could not be fully depicted. 
The $6.37\,\mathrm{Hz}$ QPO found in $2.07\text{--}3.25\,\mathrm{s}$ is our focus.
To ensure adequate frequency resolution, 
we select the $100 \text{--}300\,\mathrm{keV}$ emission in the time interval of $2.07 \text{--} 22.07\,\mathrm{s}$ ($\Delta{f} = 0.05\,\mathrm{Hz}$) rather than
in the time interval of $2.07 \text{--} 3.25\,\mathrm{s}$
to perform the Fourier analysis.
We note that if the large pulse before $2.07\,\mathrm{s}$ is involved in the Fourier analysis, an extraneous red noise would be introduced.
Then, the $100 \text{--} 300\,\mathrm{keV}$ emission before $2.07\,\mathrm{s}$ is not involved.
In Extended Data Fig.~\ref{Fig_FP_GLSP_compare},
we show the Fourier periodograms
of $100 \text{--} 300\,\mathrm{keV}$ emission during $2.07\text{--}3.25\,\mathrm{s}$ and $2.07 \text{--} 22.07\,\mathrm{s}$ with the green and blue lines, respectively.
For the blue line, the frequency resolution is $\Delta{f} = 0.05\,\mathrm{Hz}$ and thus can well depict the overall characteristics of $6.37\,\mathrm{Hz}$ QPO.
This is evident in Extended Data Fig.~\ref{Fig_FP_GLSP_compare}.
According to the blue line,
it seems that the power of QPOs dominates the power of periodograms in $f \sim 2\text{--}10\,\mathrm{Hz}$.
We would also like to point out that
the ZK-normalized power (orange line in Extended Data Fig.~\ref{Fig_FP_GLSP_compare}) based on GLSP analysis for the $100 \text{--} 300\,\mathrm{keV}$ emission during $2.07\text{--}3.25\,\mathrm{s}$ is well consistent with the Fourier periodograms of $100 \text{--} 300\,\mathrm{keV}$ emission in $2.07 \text{--} 22.07\,\mathrm{s}$.
It reveals that GLSP analysis well reveals the overall characteristics of low-frequency QPOs.

\textbf{Modeling Broadband Noise and QPOs.\;}
Two empirical models are generally adopted to model the broadband variability. 
The first model is a simple power-law (PL) model with a constant term for white noise, i.e.,
\begin{equation}
\label{eq:fft_pl}
	P_{\rm bn}(f) = A f^{-\beta} + C,
\end{equation}
where $A$ is the normalization factor, 
$\beta$ is the power-law index,
$f$ is the sampling frequency bounded between $1 / T$ (inverse of the total time span) and the Nyquist frequency $1 / (2 \cdot \delta{t})$ (with $\delta{t}$ being the time bin size), 
and $C$ represents the white noise level.
The second model is a bending power-law (BPL) model with a constant term for white noise, i.e.,
\begin{equation}
\label{eq:fft_bpl}
	P_{\rm bn}(f) = \frac{A f^{-\alpha}}{1 + (f / f_{\mathrm{bend}})^{\beta - \alpha}} + C,
\end{equation}
where $\alpha$, $\beta$ and $f_{\mathrm{bend}}$ are the low-frequency slope, 
high-frequency slope, and the characteristic bended frequency, respectively. 
The profile of an QPO can be generally described with a Lorentzian function. 
In a scenario involving $n$ QPOs, 
the power in the periodogram can be described as: 
\begin{equation}
\label{eq:fft_ltzs}
	P_{\rm QPOs}(f) = \sum_{i=1}^{n} \frac{A_{i} \times w_{i}}{\pi [(f - f_{0,i})^{2} + w_{i}^{2}]},
\end{equation}
where $f_{0, i}$ is the centroid frequency of the $i$-th QPO, 
$w_{i}$ is the half-width of the Lorentzian component, and $A_{i}$ represents the integrated power under the $i$-th Lorentzian.

We conducted parameter estimation using the nested sampler Dynesty\cite{2004AIPC..735..395S,10.1214/06-BA127,2020MNRAS.493.3132S,sergey_koposov_2024_12537467} (implemented via Bilby\cite{bilby_paper}).
In our maximum likelihood estimation, 
we employ the Whittle likelihood function ${\mathcal{L}}$ \cite{10.1007/BF02590998},
\begin{equation}
\label{eq:fft_lnL}
    \ln{\mathcal{L}} = -\sum \left( \frac{P}{P_{\rm m}} + \log{P_{\rm m}}  \right),
\end{equation}
where $P$ denotes the observed power spectral density and $P_{\rm m}$ represents the model-predicted spectral power.

\begin{figure}
	\centering
	\includegraphics[width=1.0\textwidth]{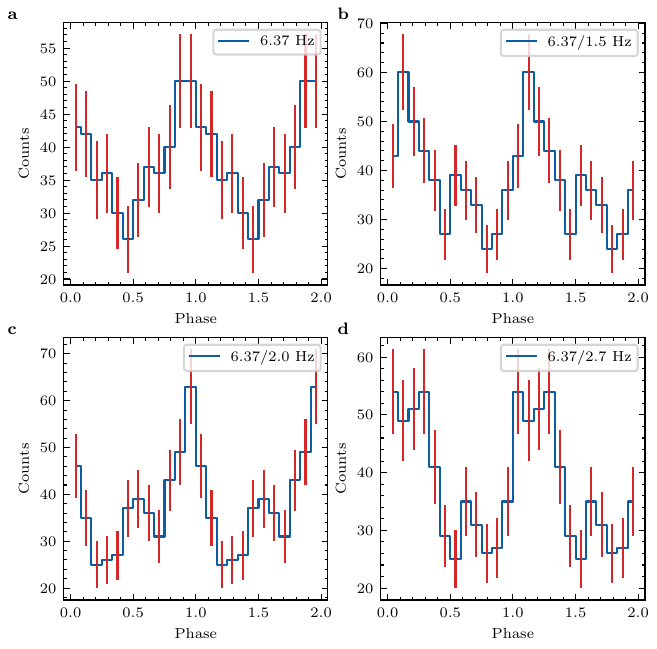}
	\caption{Phase-folded Profiles of BAT $100 \text{--} 300\,\mathrm{keV}$. 
    (\textbf{a}). Phase-folded at $6.37\,\mathrm{Hz}$, 
    displaying a single-peaked profile. 
    (\textbf{b}). Phase-folded at $6.37 / 1.5\,\mathrm{Hz}$, 
    showing a double-peaked structure with a dominant peak and a secondary sub-peak. 
    (\textbf{c}). Phase-folded at $6.37 / 2.0\,\mathrm{Hz}$, 
    exhibiting a similar double-peaked morphology. 
    (\textbf{d}). Phase-folded at $6.37 / 2.7\,\mathrm{Hz}$, 
    with complex cyclic morphology.}
	\label{Fig_fold}
\end{figure}

\begin{figure}
	\centering
	\includegraphics[width=0.48\textwidth]{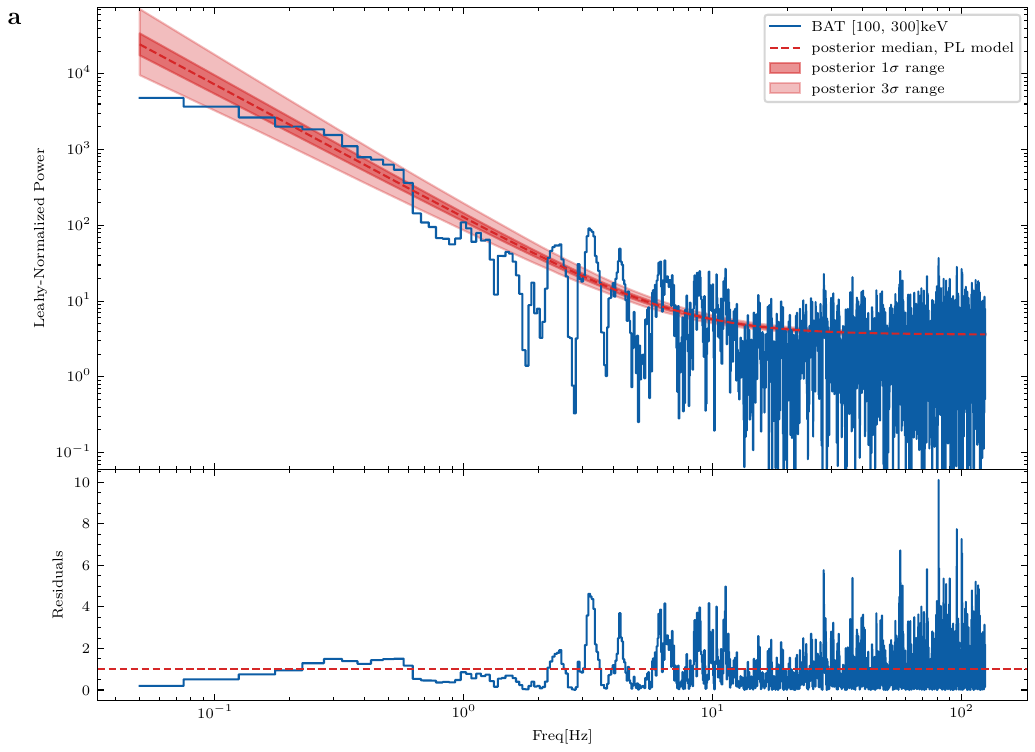}
	\includegraphics[width=0.48\textwidth]{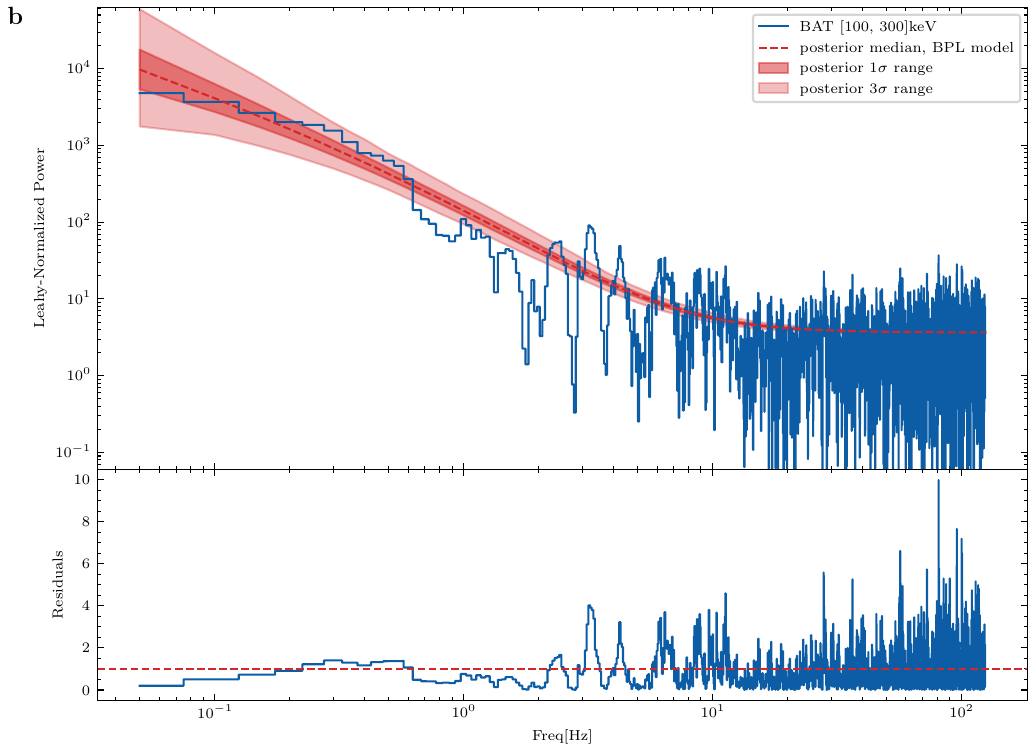}
	\includegraphics[width=0.48\textwidth]{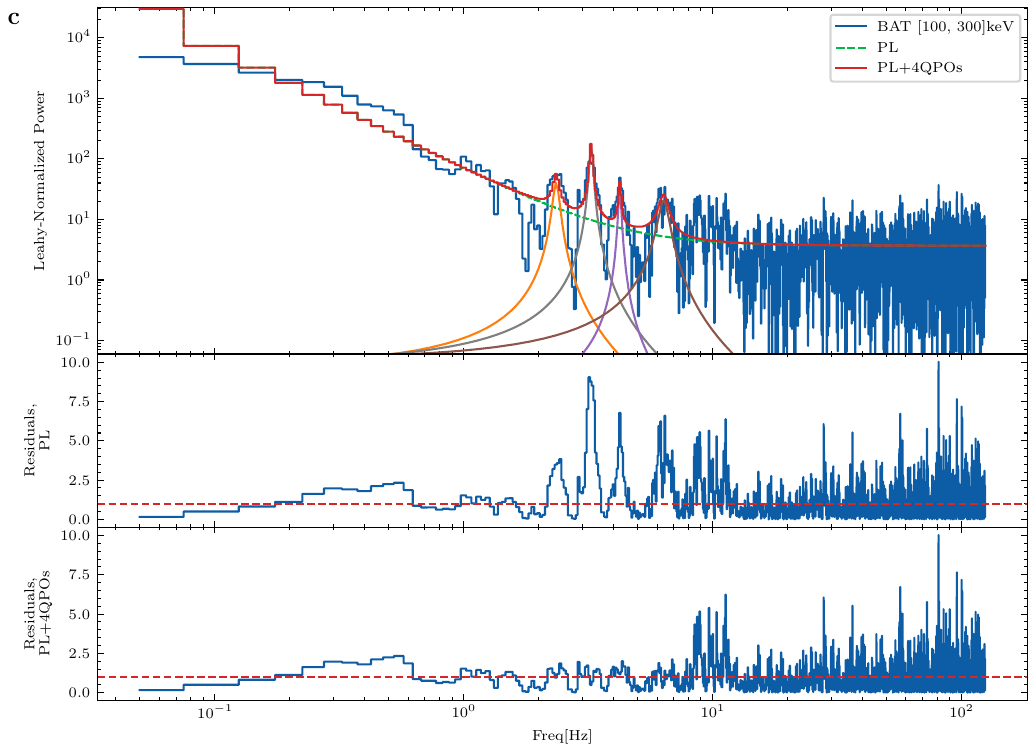}
	\includegraphics[width=0.48\textwidth]{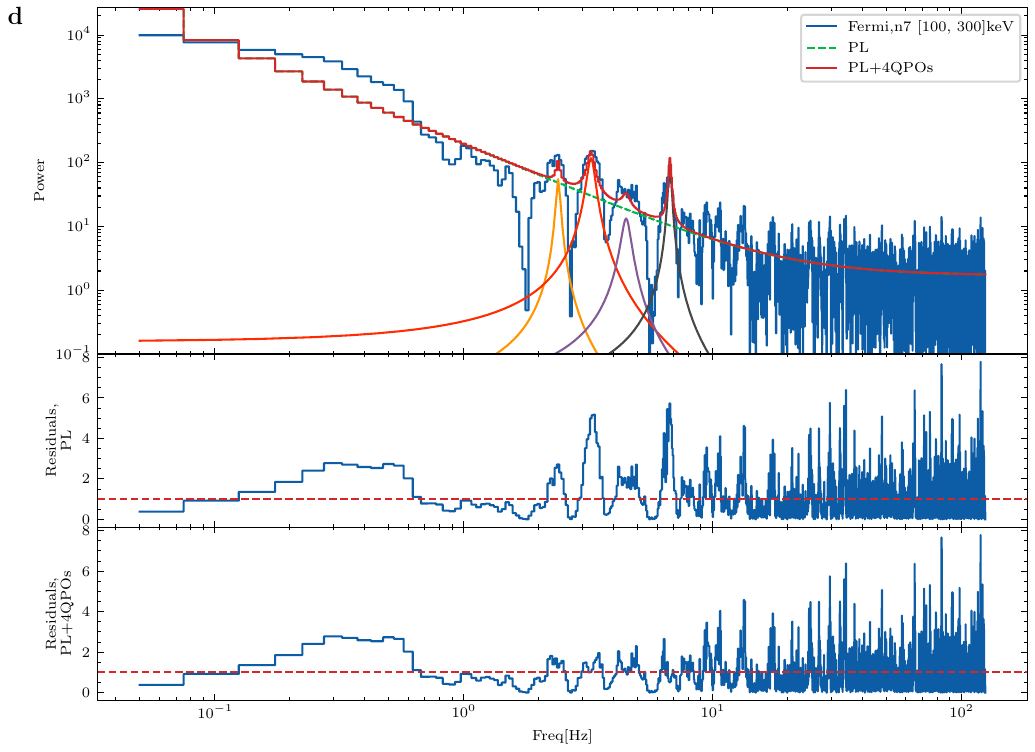}
	\caption{The fitting results of the periodogram for the $100 \text{--} 300\,{\rm keV}$ light curve in the $2.07 \text{--} 22.07\,{\rm s}$ time interval are presented. 
    Each panel is for a different model: 
    (\textbf{a}) PL model, 
    (\textbf{b}) BPL model, 
    (\textbf{c}) PL$+$4QPOs model, 
    (\textbf{d}) PL$+$4QPOs model for Fermi data.}
	\label{Fig_FFT_100to300}
\end{figure}

\begin{figure}
	\centering
        \includegraphics[width=0.48\textwidth]{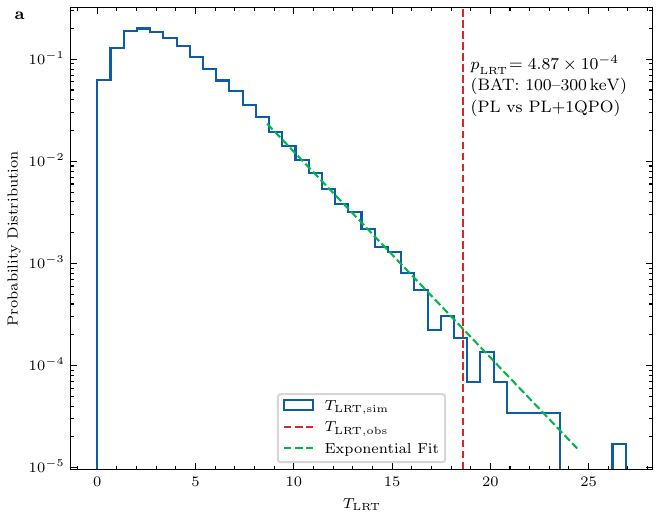}
        \includegraphics[width=0.48\textwidth]{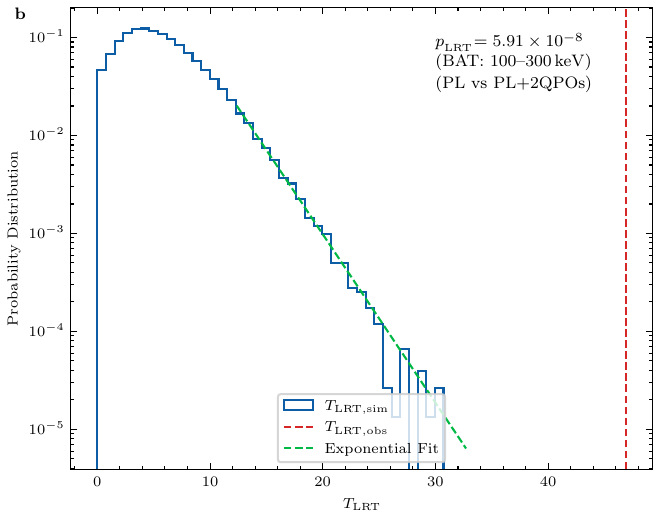}
	\includegraphics[width=0.48\textwidth]{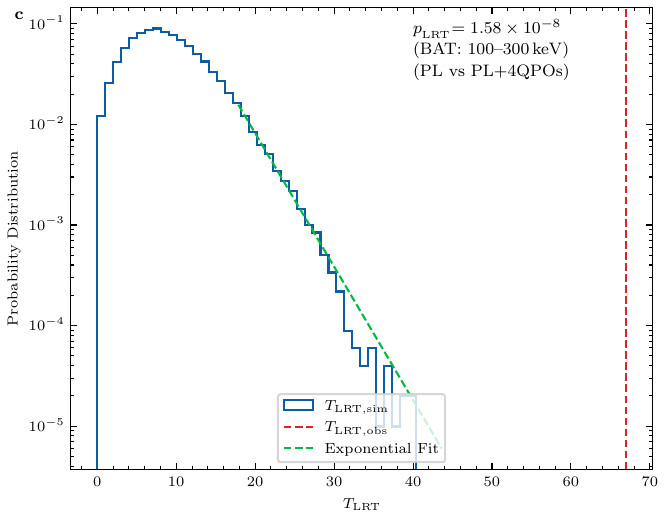}
	\includegraphics[width=0.48\textwidth]{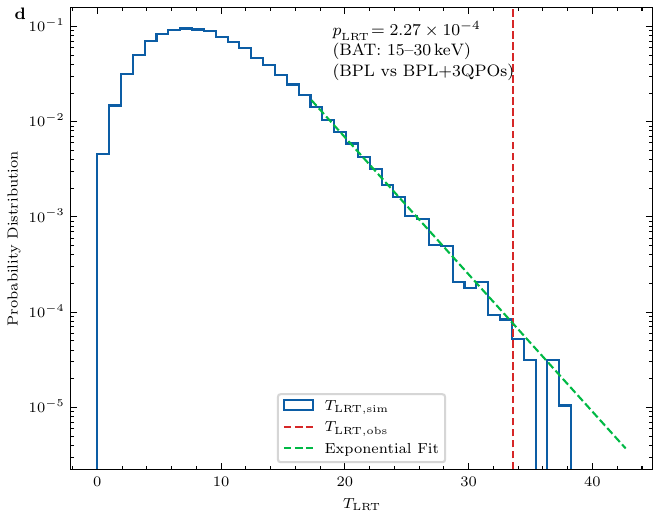}
	\caption{The probability distributions of $T_{\rm LRT}$, 
    derived from simulated periodograms under the null hypothesis.
    Vertical red dashed lines mark the observed $T_{\rm LRT}$ in Swift data.
    Green dashed lines indicate an exponential fit to the tail of the probability distribution. 
    The inset values provide the estimated false positive probability for signals stronger than the observed one. 
    Probability distributions of $T_{\rm LRT}$ under null vs. alternative hypotheses: 
    (\textbf{a}) PL vs PL$+$1QPO; 
    (\textbf{b}) PL vs PL$+$2QPOs;
    (\textbf{c}) PL vs PL$+$4QPOs;
    (\textbf{d}) BPL vs BPL$+$3QPOs.
	}
	\label{Fig_TLR}
\end{figure}

\textbf{Model selection for broadband noise and Signal Significance Testing.\;}
In the following, we model the periodogram data.
\begin{itemize}
\item \textit{Broadband noise modeling: BPL vs. PL models}.\;
We calculate the likelihood ratio $T_{\rm TRT,obs}$
between the fitting result based on the BPL model
and that based on the PL model.
To assess whether the fitting with the BPL model provides a statistically significant improvement over the fitting with the PL model,
a standard test statistic for nested models based on the likelihood ratio is adopted
(see ref.~\cite{2010MNRAS.402..307V} for methodological details).
The procedure is as follows.
In the following, the null hypothesis ($H_{0}$) is taken as the PL model and the alternative hypothesis ($H_{1}$) is the BPL model. 
A number of $10^{5}$ periodograms according to the posterior distribution of fittings based on the PL model are produced,
and fitted with both the PL model and the BPL model,
respectively.
Then, one can obtain the value of the likelihood ratio
$T_{\rm LRT, sim}$.
Based on the $T_{\rm LRT, sim}$ distribution
and the value of the observed $T_{\rm LRT, obs}$,
one can obtain the probability of 
\begin{equation}\label{eq:fft_LRT}
	p_{_{\rm LRT}}= {\rm Pr}[T_{{\rm LRT, sim}} > T_{{\rm LRT, obs}}]. 
\end{equation}

A small value of $p_{_{\rm LRT}}$ indicates that the data are less likely to be drawn from the null hypothesis $H_0$, 
implying that the complex model $H_1$ may represent the data better than the null hypothesis.
We would like to choose the BPL model rather than the PL model if $p_{_{\rm LRT}} < 10^{-2}$ is obtained.
The identified optimal model for the broadband variability can be found in the last column of the Extended Data Table~\ref{exttab3}.
It reveals that the BPL model provides no statistically significant improvement over the PL model for characterizing broadband variability of the $100\text{--}300\,\mathrm{keV}$ emission's periodogram.
Therefore, we adopt the PL model in the following discussions.

\item \textit{QPOs identified}.\;
Based on the identified optimal model for the broadband variability,
one can find several distinct QPOs in the residuals of the periodogram.
For example, there seems to be four distinct QPOs in the residuals of the periodogram for the $100\text{--}300\,\mathrm{keV}$ during $2.07\text{--}22.07\,\mathrm{s}$, 
as shown in the bottom panel of Extended Data Fig.~\ref{Fig_FFT_100to300}\textbf{a-b}.
Besides the $6.37\,\mathrm{Hz}$ QPO, which is also identified via WWZ and GLSP analysis,
there seems to be three QPOs at $6.37/2.7\,\mathrm{Hz}$, 
$6.37/2\,\mathrm{Hz}$, and $6.37/1.5\,\mathrm{Hz}$.
In order to illustrate, the phase-folded profiles at these frequencies for the $100 \text{--} 300\,\mathrm{keV}$ emission,
generated based on Stingray \textit{fold\_events},
are shown in Extended Data Fig.~\ref{Fig_fold}. 
\begin{itemize}
\item
Phase-folded profile at $6.37\,\mathrm{Hz}$ (Extended Data Fig.~\ref{Fig_fold}\textbf{a}) presents a perfect characteristic curve of a trigonometric function.
This kind of profile reveals a stable harmonic modulation at this frequency.
\item
Phase-folded profile at $6.37/1.5\,\mathrm{Hz}$ (Extended Data Fig.~\ref{Fig_fold}\textbf{b}) exhibits a double-peaked structure with a primary peak at phase $0.1$ and a secondary peak at phase $0.55$.
\item
Similarly, phase-folded profile at $6.37/2\,\mathrm{Hz}$ (Extended Data Fig.~\ref{Fig_fold}\textbf{c}) is a bimodal profile combining primary peak at phase of $0.9$ and secondary peak at phase of $0.55$.
\item
Phase-folded profile at $6.37/2.7\,\mathrm{Hz}$ (Extended Data Fig.~\ref{Fig_fold}\textbf{d}) exhibits a complex cyclic morphology.
\end{itemize}
The multipeaked profile
obtained based on the phase-folded method,
e.g., the phase-folded profile at $6.37/1.5\,\mathrm{Hz}$ or $6.37/2.7\,\mathrm{Hz}$,
may not be associated with independent oscillations relative to the $6.37\,\mathrm{Hz}$ QPO.
The $6.37/2.0\,\mathrm{Hz}$ QPO may be the subharmonic of the $6.37\,\mathrm{Hz}$ QPO.
However, 
it remains unknown whether the $6.37/1.5\,\mathrm{Hz}$ and $6.37/2.7\,\mathrm{Hz}$ signals are associated with the $6.37\,\mathrm{Hz}$ oscillation.
Then, we fit the $100\text{--}300\,\mathrm{keV}$ periodogram
based on the model of
PL$+$1QPO (for $6.37\,\mathrm{Hz}$),
PL$+$2QPOs (for $6.37\,\mathrm{Hz}$ and $6.37/2\,\mathrm{Hz}$),
and PL$+$4QPOs (for $6.37\,\mathrm{Hz}$, $6.37/1.5\,\mathrm{Hz}$, $6.37/2\,\mathrm{Hz}$, and $6.37/2.7\,\mathrm{Hz}$), respectively.
The fitting results of the periodograms for the $100 \text{--} 300\,\mathrm{keV}$ energy bands are presented in Extended Data Fig.~\ref{Fig_FFT_100to300}.

\item \textit{QPOs' significance estimation}.\;
The requirement of the QPO component in the above models is estimated statistically by comparing the fitting result with the model of PL.
The value of likelihood ratio $T_{\rm LRT}$ for models with and without QPOs for the observations ($T_{\rm LRT,obs}$) and the simulation data ($T_{\rm LRT,sim}$) are all calculated,
where the null model is that without the QPO component.
The corresponding $p_{_{\rm LRT}}={\rm Pr}[T_{\rm LRT,sim}>T_{\rm LRT,obs}]$ is estimated and used to assess their statistical significance. 
The probability distribution of $T_{\rm LRT,sim}$ is showed in Extended Data Fig.~\ref{Fig_TLR},
which exhibits a linear trend on the log-linear plot for the tail.
In order to estimate the value of $p_{_{\rm LRT}}$,
the tail of $T_{\rm LRT,sim}$ distribution from its $95\%$ to $99.9\%$ percentile
is fitted with an exponential function $a \cdot \exp{\left(- b x \right)}$, where $a$ and $b$ are constants.
The value of $p_{_{\rm LRT}}$ can be obtained with $p_{_{\rm LRT}} = \int_{T_{\rm LRT,obs}}^{\infty} a \cdot \exp{\left(- b x \right)} d{x}$.
\end{itemize}

Detailed fitting parameters corresponding to these analyses can be found in Extended Data Table~\ref{exttab3}.
The prior distributions for the parameters adopted in our analysis are tabulated in Extended Data Table~\ref{exttab4}.

\subsection*{Sinusoidal Modulation\;}\label{subsec:sin}
For the BAT $100 \text{--} 300\,{\rm keV}$ light curve with $16\,{\rm ms}$ time binning, 
we applied a Butterworth low-pass filter to obtain the low-pass component (green dashed curve in Fig.~\ref{Fig_multi_lc}\textbf{b}). 
We then model the residual light curve 
(subtracting its low-pass component from the original; 
pink data points in Fig.~\ref{Fig_multi_lc}\textbf{c}) within the $2.07 \text{--} 3.25\,{\rm s}$ interval using a sinusoidal function $f(t) = A_{\rm s} \sin{(2\pi f_{\rm s} t + \phi_{\rm s})}$.
The optimal result of our fitting is obtained as $A_{\rm s} = 0.32^{+0.04}_{-0.04}$,
$f_{\rm s} = 6.33^{+0.05}_{-0.05}\,{\rm Hz}$,
and $\phi_{\rm s} = 3.79^{+0.82}_{-0.89}\,{\rm rad}$.

\subsection*{QPO Detection in Non-thermal Components\;}\label{subsec:non-thermal}

\begin{figure}
	\centering
	\includegraphics[width=0.45\textwidth]{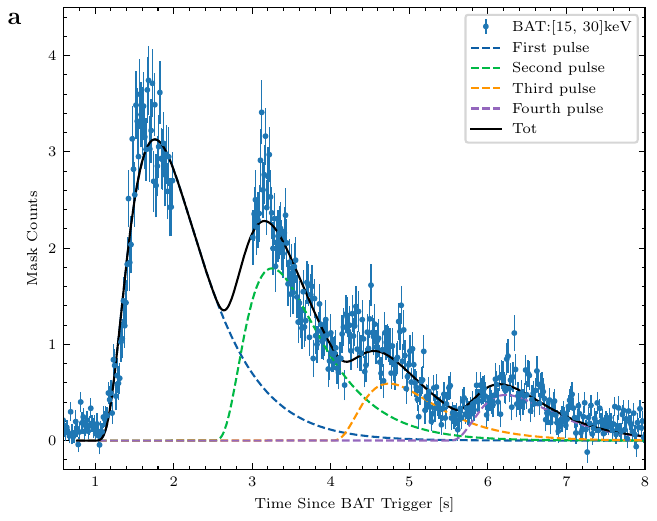}
    \includegraphics[width=0.45\textwidth]{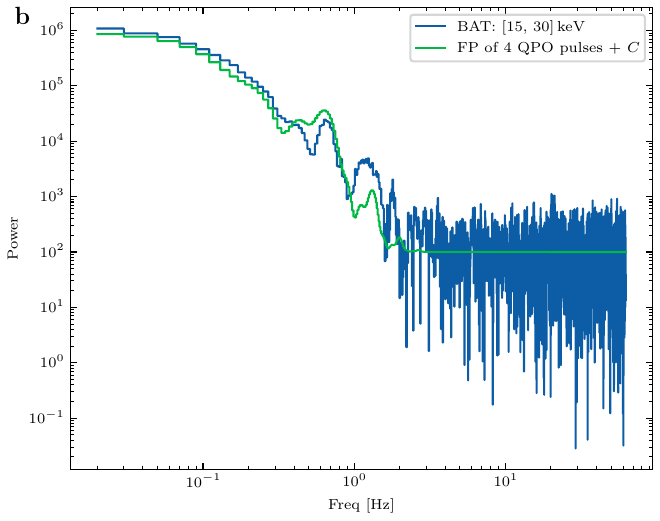}
	\caption{(\textbf{a}) BAT $15 \text{--} 30\,\mathrm{keV}$ light curve excluding the $2\text{--}3\,\mathrm{s}$ interval. 
    Black solid line shows the best-fit periodic FRED model for four pulses; 
    dashed curves indicate individual FRED components. 
    (\textbf{b}) The Fourier periodogram (FP) of the original light curve and that of the fitted periodic FRED pulses. 
Both the low-frequency broadband noise and QPO features are consistent for those from the observation and the periodic FRED pulses model.}
	\label{Fig_syn_lc_fit}
\end{figure}

\textbf{Non-thermal Pulses Fitting.\;}
Several functional forms have been proposed to fit the pulse shapes of GRB.
The fast-rising exponential decay (FRED) model is ubiquitous in GRB pulse modeling.
We employ the following FRED model\cite{1996ApJ...459..393N,2021NatAs...5..560P} to fit the light-curve of a non-thermal emission pulse,
\begin{equation}
\label{eq:FRED}
    I(t|A, \tau, \zeta, \Delta) = A \exp{[-\zeta (\frac{t - t_{\rm max}}{\tau} + \frac{\tau}{t - \Delta})]}
\end{equation}
where $A$ is the normalization parameter,
$\tau$ is a duration scaling parameter,
$\Delta$ describes the time delay of the pulse with respect to the trigger time,
and $\zeta$ is an asymmetry parameter
used to adjust the skewness of the pulse.
The periodic FRED pulses are our focus and are described as follows:
\begin{equation}
\label{eq:FRED_qpo}
    S_{\rm sup} = \sum_{i=1}^{n} I(t | A_{i}, \tau, \zeta, \Delta + (i - 1)\delta_{\rm t}),
\end{equation}
where $i$ represents the $i$-th independent FRED pulse,
$n$ is the total number of pulses,
$\delta_{t}$ is the period time,
and $A_{i}$ is the amplitude of each pulse.
In this equation, the pulse's profile is the same but with different amplitude and the beginning time.
The data during $2 \text{--} 3\,{\rm s}$,
of which the thermal emission makes a significant contribution,
is not involved in our fitting. 
Please see Extended Data Fig.~\ref{Fig_syn_lc_fit}\textbf{a}.
The fitting result is shown with a black line in Extended Data Fig.~\ref{Fig_syn_lc_fit}\textbf{a}, 
while the blue, green, orange, and purple dashed-lines represent the first, second, third, and fourth pulses, respectively.
Posterior distributions of the model parameters are shown in the Extended Data Table.~\ref{exttab5}, 
with a period of $\delta_{t} = 1.49\,\mathrm{s}$ (corresponding to a frequency of $\mathord{\sim}0.67\,\mathrm{Hz}$).

Since the thermal emission component makes a significant contribution to the $15 \text{--} 30\,\mathrm{keV}$ emission during $2 \text{--} 3\,{\rm s}$,
we plot the modified data (red dots) in Fig.~\ref{Fig_LSP_syn}\textbf{a}.
The modified data is the data with thermal emission subtracted and is obtained as follows.
The time interval of $2.07 \text{--} 3.25\,{\rm s}$ is divided it into 3 time slices,
i.e.,
$2.07 \text{--} 2.47\,\mathrm{s}$, $2.47 \text{--} 2.87\,\mathrm{s}$, $2.87 \text{--} 3.27\,\mathrm{s}$.
We used 2SBPL$+$CPL-mBB model to fit the radiation spectrum in these 3 time slices.
Based on spectral fitting results,
the contribution proportion of the non-thermal emission to the $15 \text{--} 30\,\mathrm{keV}$ flux is obtained as
$0.84^{+0.02}_{-0.01}$, $0.80^{+0.01}_{-0.01}$, and $0.88^{+0.02}_{-0.02}$ for 3 time slices, respectively.
The contribution proportion factor in each time slice is used to modify the light-curve
by multiplying the factor on the flux of $15 \text{--} 30\,\mathrm{keV}$.
The result is showed in Fig.~\ref{Fig_LSP_syn}\textbf{a} with red dots.
The modified data are well consistent with the model light-curve.

\textbf{Fourier analysis on the non-thermal emission.\;}
It is found that the non-thermal emission may exhibit a $\mathord{\sim}0.67\,\mathrm{Hz}$ QPO.
Then, we perform the Fourier analysis on the non-thermal emission to identify the corresponding QPO component.
Since a longer time interval can provide a better frequency resolution, 
we select the $15\text{--}30\,\mathrm{keV}$ emission from $-10$~s to $40$~s ($\Delta{f} = 0.02\,\mathrm{Hz}$) for our Fourier analysis.

\begin{figure}
	\centering
	\includegraphics[width=0.48\textwidth]{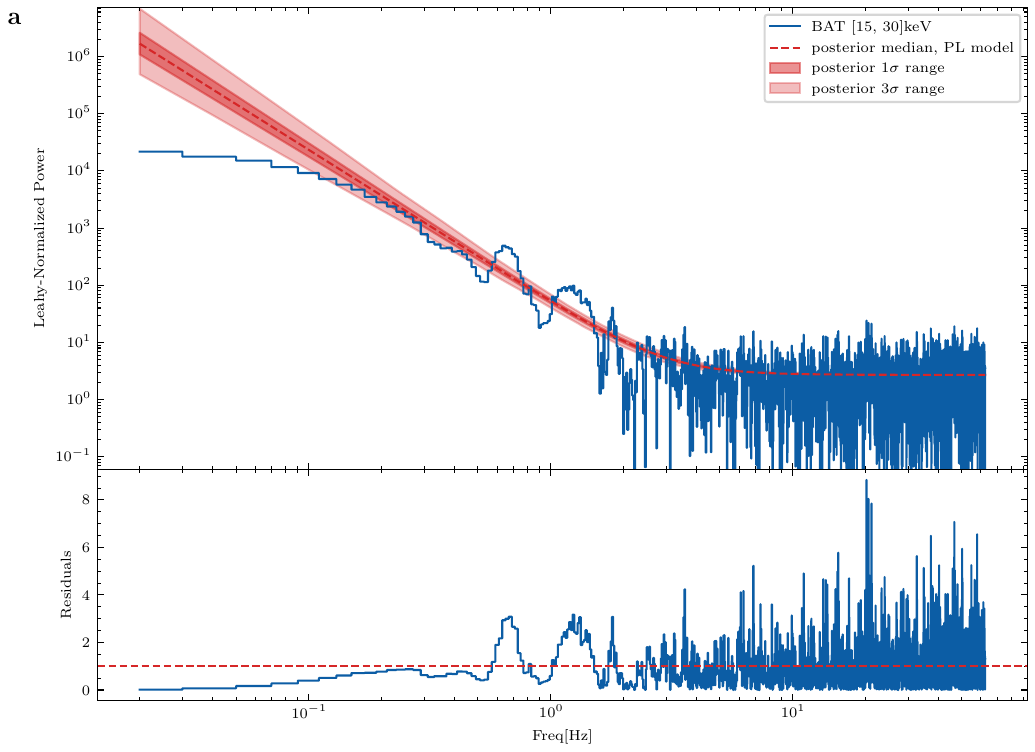}
	\includegraphics[width=0.48\textwidth]{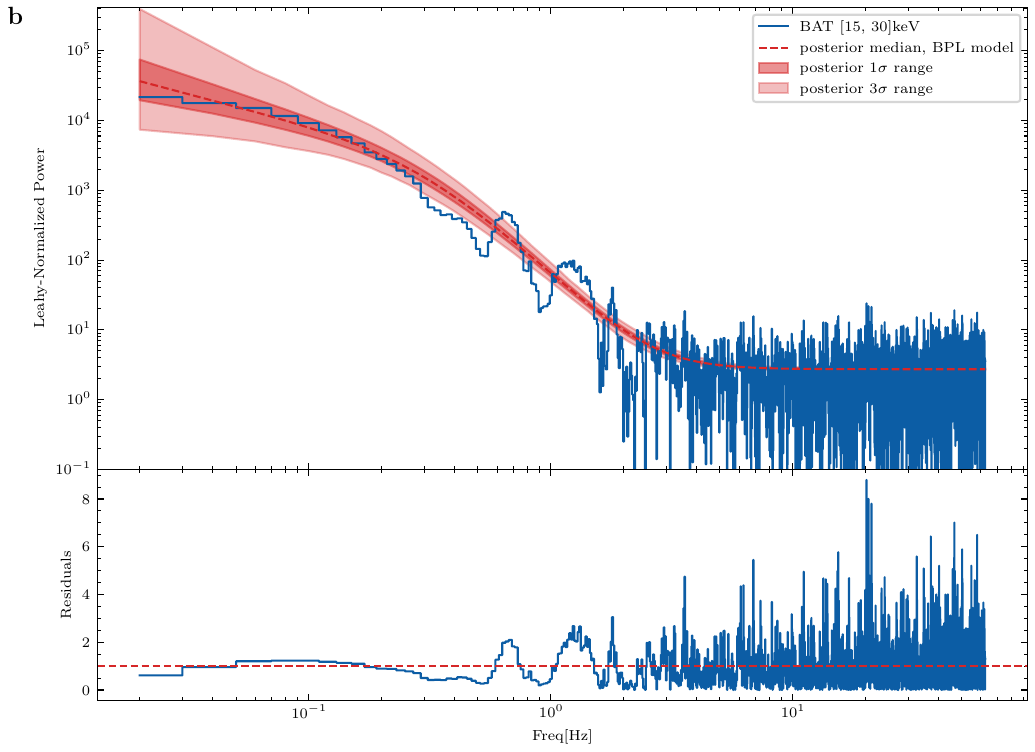}
	\includegraphics[width=0.48\textwidth]{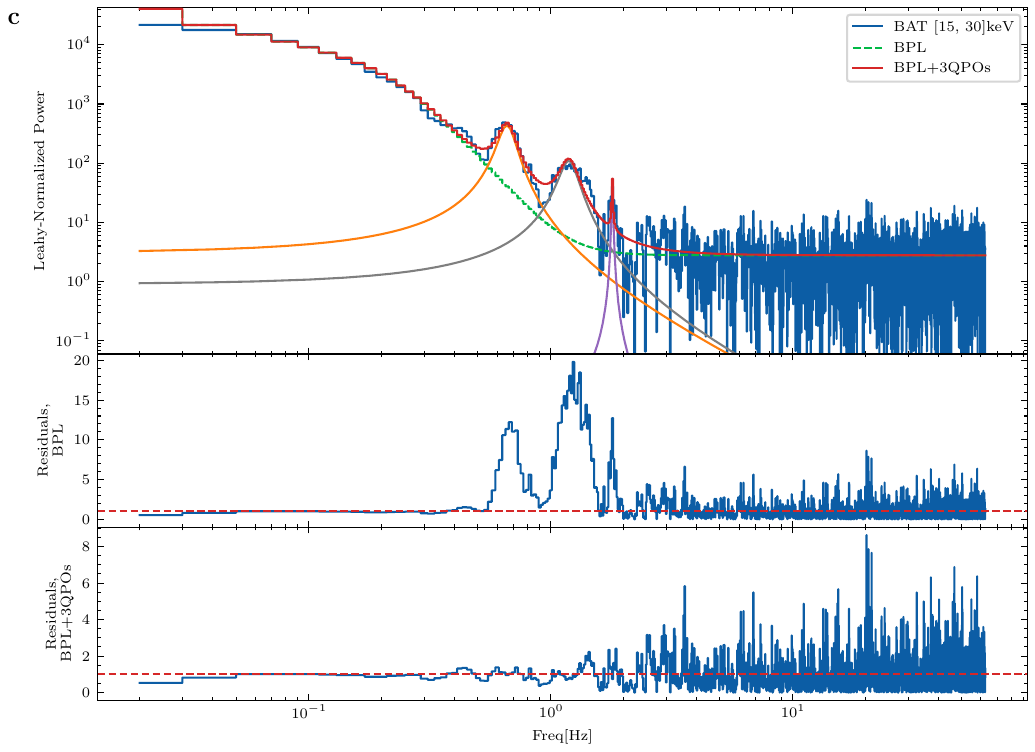}
	\includegraphics[width=0.48\textwidth]{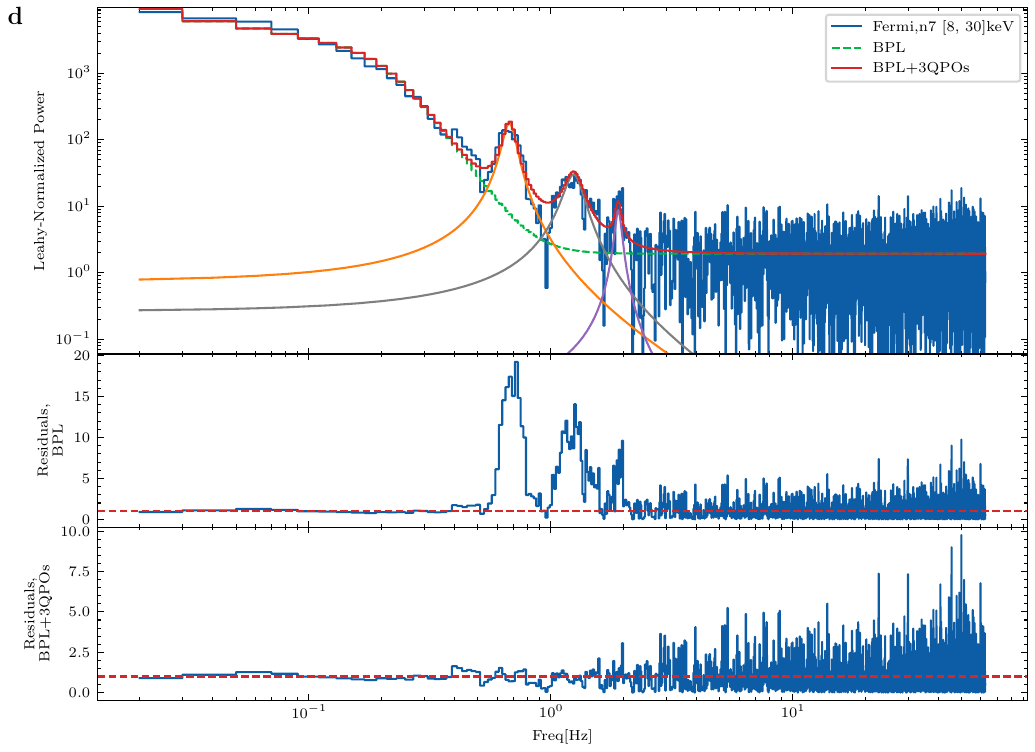}
	\caption{The periodgram fitting results for the light curves in the time interval $[-10, 40]\,{\rm s}$ and energy bands of $15 \text{--} 30\,{\rm keV}$ ($8 \text{--} 30\,{\rm keV}$ for Fermi) are presented.
    Each panel is for a different model: 
    (\textbf{a}) PL model, 
    (\textbf{b}) BPL model, 
    (\textbf{c}) BPL$+$3QPOs model, 
    (\textbf{d}) BPL$+$3QPOs model for Fermi $8 \text{--} 30\,{\rm keV}$ data.}
	\label{Fig_FFT_15to30}
\end{figure}

Based on the likelihood ratio tests,
we first select the optimal model for the broadband variability of the $15\text{--}30\,\mathrm{keV}$ emission,
i.e., BPL (see Extended Data Table~\ref{exttab3}).
As showed in Extended Data Fig.~\ref{Fig_FFT_15to30}\textbf{b},
the residuals from the BPL model fitting reveal three distinct QPOs,
i.e., $\sim 0.63\,\mathrm{Hz}$, $0.63 \times 2\,\mathrm{Hz}$,
and $0.63 \times 3\,\mathrm{Hz}$ QPOs.
The $0.63 \times 2\,\mathrm{Hz}$ and $0.63 \times 3\,\mathrm{Hz}$ components are harmonics of the $0.63\,\mathrm{Hz}$ signal.
We also perform the Fourier analysis on our fitting result of the light-curve with periodic FRED pulses,
and the obtained periodogram and the corresponding QPOs are all consistent with those shown in Extended Data Fig.~\ref{Fig_syn_lc_fit}\textbf{b}.
It reveals that $\mathord{\sim} 0.63\,\mathrm{Hz}$, $0.63 \times 2\,\mathrm{Hz}$, and $0.63 \times 3\,\mathrm{Hz}$ QPOs are coherent.
We therefore model the $15\text{--}30\,\mathrm{keV}$ emission's periodogram with BPL$+$3QPOs.
The fitting results of the periodograms for the $15 \text{--} 30\,\mathrm{keV}$ energy bands are presented in Extended Data Fig.~\ref{Fig_FFT_15to30}.
We also estimate the significance of the appearance of the QPO components in the periodogram.
Detailed fitting parameters corresponding to these analyses can be found in Extended Data Table~\ref{exttab3}.

\clearpage

\clearpage
\begin{table*}
    \centering
    \renewcommand{\arraystretch}{2}
    \caption{Fitting results of different spectral models.}
    \label{exttab1}
    \resizebox{\textwidth}{!}{
    \begin{tabular}{|c|c|cccc|cccccc|cc|ccc|cc|c|c|c|c|c|} 
    \hline
    \hline
        Time interval & Model & \multicolumn{4}{|c|}{Band Parameters} & \multicolumn{6}{|c|}{2SBPL Parameters} & \multicolumn{2}{|c|}{CPL-mBB parameters} & \multicolumn{3}{|c|}{CPL parameters} & $\log_{10}({\frac{F_{\rm 15-30, Band}}{F_{\rm 15-30, CPL}}})$  & $\log_{10}({\frac{F_{\rm 100-300, Band}}{F_{\rm 100-300, CPL}}})$ & PG-stat/dof & AIC & BIC & $\Delta{\rm AIC}$ & $\Delta{\rm BIC}$\\
        ~ & ~ & $\alpha$ & $\beta$ & $\log_{10}(E_{\mathrm{c}} / \text{keV})$ & Flux ($\rm erg/cm^{2}/s$) & $\alpha_{1}$ & $\alpha_{2}$ & $\beta$ & $\log_{10}(E_{\mathrm{b}} / \text{keV})$ & $\log_{10}(E_{\mathrm{p}} / \text{keV})$ & Flux ($\rm erg/cm^{2}/s$) & $\mathrm{k} T_{\mathrm{c}}$ (keV) & Flux ($\rm erg/cm^{2}/s$) & $\alpha$ & $\log_{10}{(E_{\rm c} / \text{keV})}$ & Flux ($\rm erg/cm^{2}/s$) &  $\log_{10}({\frac{F_{\rm 15-30, 2SBPL}}{F_{\rm 15-30, CPL-mBB}}})$ & $\log_{10}({\frac{F_{\rm 100-300, 2SBPL}}{F_{\rm 100-300, CPL-mBB}}})$ & ~ & ~ & ~ & ~ & ~ \\ \hline
        \multirow{4}{*}{$1.00\text{--}1.73\,\mathrm{s}$} & Band & $-0.36^{+0.02}_{-0.02}$ & $-2.38^{+0.04}_{-0.04}$ & $2.44^{+0.02}_{-0.02}$ & $6.58^{+0.12}_{-0.12} \times 10^{-5}$ & -- & -- & -- & -- & -- & -- & -- & -- & -- & -- & -- & -- & -- & $388.76/329$ & 396.76 & 411.99 & 28.80 & 13.57\\ 
        ~ & 2SBPL & -- & -- & -- & -- & $-0.44^{+0.03}_{-0.03}$ & $-1.38^{+0.08}_{-0.08}$ & $-2.61^{+0.07}_{-0.08}$ & $2.10^{+0.04}_{-0.04}$ & $2.80^{+0.04}_{-0.03}$ & $6.54^{+0.12}_{-0.13} \times 10^{-5}$ & -- & -- & -- & -- & -- & -- & -- & $353.93/327$ & 365.93 & 388.78 & -2.03 & -9.64\\ 
        ~ & Band + CPL & $-0.26^{+0.08}_{-0.05}$ & $-2.36^{+0.09}_{-0.09}$ & $2.31^{+0.05}_{-0.06}$ & $4.75^{+0.48}_{-0.56} \times 10^{-5}$ & -- & -- & -- & -- & -- & -- & -- & -- & $0.31^{+0.29}_{-0.47}$ & $2.82^{+0.11}_{-0.09}$ & $1.77^{+0.52}_{-0.43} \times 10^{-5}$ & $1.05^{+0.71}_{-0.57}$ & $0.82^{+0.34}_{-0.30}$ & $358.36/326$ & 372.36 & 399.01 & 4.40 & 0.59 \\ 
        ~ & 2SBPL + CPL-mBB & -- & -- & -- & -- & $-0.60^{+0.05}_{-0.07}$ & $-1.57^{+0.17}_{-0.20}$ & $-2.88^{+0.19}_{-0.30}$ & $2.09^{+0.06}_{-0.08}$ & $3.17^{+0.14}_{-0.10}$ & $4.37^{+0.45}_{-0.42} \times 10^{-5}$ & $30.68^{+7.50}_{-4.58}$ & $2.26^{+0.39}_{-0.45} \times 10^{-5}$ & -- & -- & -- & $0.54^{+0.22}_{-0.23}$ & $0.02^{+0.16}_{-0.19}$ & $351.96/325$ & 367.96 & 358.36 & -- & --\\
        \hline
        \multirow{4}{*}{$1.73\text{--}2.07\,\mathrm{s}$} & Band & $-0.39^{+0.04}_{-0.04}$ & $-2.15^{+0.02}_{-0.02}$ & $2.20^{+0.03}_{-0.03}$ & $6.50^{+0.14}_{-0.12} \times 10^{-5}$ & -- & -- & -- & -- & -- & -- & -- & -- & -- & -- & -- & -- & -- & $411.45/359$ & 419.45 & 435.03 & 29.88 & 14.30 \\ 
        ~ & 2SBPL & -- & -- & -- & -- & $-0.63^{+0.03}_{-0.03}$ & $-1.99^{+0.01}_{-0.01}$ & $-2.97^{+0.16}_{-0.23}$ & $2.09^{+0.01}_{-0.01}$ & $4.19^{+0.19}_{-0.13}$ & $7.14^{+0.12}_{-0.10} \times 10^{-5}$ & -- & -- & -- & -- & -- & -- & -- & $435.43/357$ & 447.43 & 470.80 & 57.86 & 50.07 \\ 
        ~ & Band + CPL & $-0.39^{+0.04}_{-0.04}$ & $-2.17^{+0.03}_{-0.05}$ & $2.21^{+0.03}_{-0.03}$ & $6.42^{+0.16}_{-0.23} \times 10^{-5}$ & -- & -- & -- & -- & -- & -- & -- & -- & $1.21^{+0.50}_{-0.50}$ & $4.83^{+0.55}_{-0.59}$ & $2.03^{+0.18}_{-0.02} \times 10^{-7}$ & $3.50^{+1.47}_{-1.46}$ & $3.58^{+1.80}_{-1.20}$ & $404.96/356$ & 418.96 & 446.22 & 29.39 & 25.49 \\ 
        ~ & 2SBPL + CPL-mBB & -- & -- & -- & -- & $-0.68^{+0.05}_{-0.05}$ & $-1.92^{+0.06}_{-0.05}$ & $-2.85^{+0.28}_{-0.48}$ & $1.93^{+0.04}_{-0.05}$ & $4.63^{+0.28}_{-0.27}$ & $4.22^{+0.33}_{-0.30} \times 10^{-5}$ & $25.32^{+2.32}_{-1.83}$ & $2.06^{+0.23}_{-0.23} \times 10^{-5}$ & -- & -- & -- & $0.69^{+0.12}_{-0.12}$ & $0.01^{+0.15}_{-0.15}$ & $373.57/355$ & 389.57 & 420.73 & -- & -- \\ 
        \hline
        \multirow{4}{*}{$2.07\text{--}3.25\,\mathrm{s}$} & Band & $-0.78^{+0.01}_{-0.01}$ & $-2.23^{+0.01}_{-0.01}$ & $2.59^{+0.02}_{-0.02}$ & $6.12^{+0.07}_{-0.07} \times 10^{-5}$ & -- & -- & -- & -- & -- & -- & -- & -- & -- & -- & -- & -- & -- & $490.09/359$ & 498.09 & 590.01 & 99.40 & 83.83\\ 
        ~ & 2SBPL & -- & -- & -- & -- & $-0.92^{+0.01}_{-0.01}$ & $-1.99^{+0.01}_{-0.01}$ & $-3.11^{+0.18}_{-0.20}$ & $2.32^{+0.01}_{-0.01}$ & $3.92^{+0.22}_{-0.25}$ & $6.79^{+0.07}_{-0.07} \times 10^{-5}$ & -- & -- & -- & -- & -- & -- & -- & $578.01/357$ & 590.01 & 613.38 & 191.32 & 183.54\\ 
        ~ & Band + CPL & $-0.78^{+0.03}_{-0.02}$ & $-2.54^{+0.09}_{-0.12}$ & $2.60^{+0.02}_{-0.02}$ & $5.11^{+0.30}_{-0.30} \times 10^{-5}$ & -- & -- & -- & -- & -- & -- & -- & -- & $1.24^{+0.12}_{-0.24}$ & $4.90^{+0.09}_{-0.13}$ & $7.22^{+0.27}_{-0.25} \times 10^{-6}$ & $1.35^{+0.67}_{-0.41}$ & $1.58^{+0.44}_{-0.27}$ & $411.74/356$ & 425.74 & 453.00 & 27.05 & 23.16\\ 
        ~ & 2SBPL + CPL-mBB & -- & -- & -- & -- & $-0.81^{+0.08}_{-0.08}$ & $-1.70^{+0.02}_{-0.02}$ & $-3.05^{+0.16}_{-0.17}$ & $1.57^{+0.05}_{-0.05}$ & $4.58^{+0.07}_{-0.08}$ & $3.11^{+0.13}_{-0.13} \times 10^{-5}$ & $30.19^{+0.72}_{-0.66}$ & $2.76^{+0.07}_{-0.07} \times 10^{-5}$ & -- & -- & -- & $0.72^{+0.03}_{-0.03}$ & $-0.37^{+0.03}_{-0.03}$ & $382.69/355$ & 398.69 & 429.84 & -- & --\\ 
        \hline
        \multirow{4}{*}{$3.25\text{--}8.00\,\mathrm{s}$} & Band & $-1.24^{+0.03}_{-0.03}$ & $-2.06^{+0.02}_{-0.02}$ & $2.52^{+0.06}_{-0.05}$ & $5.76^{+0.12}_{-0.12} \times 10^{-6}$ & -- & -- & -- & -- & -- & -- & -- & -- & -- & -- & -- & -- & -- & $448.90/359$ & 456.90 & 472.47 & 69.68 & 54.10\\ 
        ~ & 2SBPL & -- & -- & -- & -- & $-1.21^{+0.06}_{-0.05}$ & $-1.83^{+0.03}_{-0.03}$ & $-3.00^{+0.22}_{-0.25}$ & $1.80^{+0.08}_{-0.08}$ & $4.50^{+0.12}_{-0.16}$ & $7.14^{+0.28}_{-0.28} \times 10^{-6}$ & -- & -- & -- & -- & -- & -- & -- & $399.69/357$ & 411.69 & 435.06 & 24.47 & 16.69\\ 
        ~ & Band + CPL & $-1.12^{+0.11}_{-0.07}$ & $-2.15^{+0.04}_{-0.05}$ & $2.37^{+0.11}_{-0.11}$ & $3.88^{+0.46}_{-0.46} \times 10^{-6}$ & -- & -- & -- & -- & -- & -- & -- & -- & $1.49^{+0.07}_{-0.10}$ & $4.82^{+0.09}_{-0.09}$ & $2.92^{+0.63}_{-0.67} \times 10^{-6}$ & $0.59^{+0.30}_{-0.25}$ & $0.58^{+0.22}_{-0.18}$ & $403.73/356$ & 417.73 & 444.99 & 30.51 & 26.62\\ 
        ~ & 2SBPL + CPL-mBB & -- & -- & -- & -- & $-0.85^{+0.06}_{-0.08}$ & $-1.77^{+0.03}_{-0.03}$ & $-2.99^{+0.21}_{-0.25}$ & $1.39^{+0.06}_{-0.06}$ & $4.52^{+0.12}_{-0.14}$ & $5.99^{+0.37}_{-0.39} \times 10^{-6}$ & $16.05^{+2.95}_{-2.32}$ & $7.33^{+1.13}_{-1.17} \times 10^{-7}$ & -- & -- & -- & $1.24^{+0.15}_{-0.14}$ & $0.35^{+0.10}_{-0.08}$ & $371.22/355$ & 387.22 & 418.37 & -- & --\\ 
        \hline
        \hline
    \end{tabular}  
    }
    \begin{tablenotes}
    \tiny
    \item The energy fluxes are calculated between $1\,{\rm keV}$ and $10\,{\rm MeV}$.
    \item All errors represent the $1 \sigma$ uncertainties.
    \item $\Delta{\rm AIC}$ and $\Delta{\rm BIC}$ values represent differences relative to the benchmark ``2SBPL$+$CPL-mBB" model.
    \item $\Delta{\rm AIC} = \mathrm{AIC_{model}} - \mathrm{AIC_{benchmark}}$, 
    $\Delta{\rm BIC} = \mathrm{BIC_{model}} - \mathrm{BIC_{benchmark}}$
    \end{tablenotes}
\end{table*}

\clearpage
\begin{table*}
	\renewcommand\arraystretch{1}
	\centering
	\caption{False alarm rate for 6.37 Hz periodic signals}
	\label{exttab2}        
	\begin{tabular}{|c|c|c|cc|}
		\hline
		\hline 
		Time resolution & Detrending models & FAP\\
		\hline
		\multirow{4}{*}{$8\,\mathrm{ms}$}& None & $1.67 \times 10^{-7}$  \\
		& 2FRED-pulses fitting & $5.20 \times 10^{-8}$ \\ 
		& Smoothing based on spline method & $3.13 \times 10^{-7}$ \\ 
	 	& Low-pass filter & $4.78 \times 10^{-7}$ \\  
		\hline
		\multirow{4}{*}{$4\,\mathrm{ms}$}& None & $2.92 \times 10^{-10}$ \\ 
		& 2FRED-pulses fitting & $1.18 \times 10^{-9}$ \\ 
		& Smoothing based on spline method & $3.22 \times 10^{-8}$ \\
		& Low-pass filter & $7.15 \times 10^{-8}$ \\ 
		\hline
		\hline
	\end{tabular}
\end{table*}

\clearpage

\begin{table*}
	\renewcommand\arraystretch{1.5}
	\centering
	\caption{Posterior summary of the Fourier analysis}
	\label{exttab3}
	\small  
	\setlength{\tabcolsep}{3pt}  
	\resizebox{\textwidth}{!}{  
		\begin{tabular}{|c|c|c|c|c|ccccccccccccccccc|ccc|}
			\hline
			\hline 
			Detetor, & Time interval (s) & Time resolution (ms) & Energy range (keV) & Noise model & $\log_{10}{A}$ & $\alpha$ & $\beta$ &  $\log_{10}{f_{\rm bend}}$  & $\log_{10}{C}$ & $\log_{10}{A_{\rm 1}}$ &  $\log_{10}{f_{0,1}}$ & $\log_{10}{w_{1}}$ & $\log_{10}{A_{\rm 2}}$ &  $\log_{10}{f_{0,2}}$ & $\log_{10}{w_{2}}$ & $\log_{10}{A_{\rm 3}}$ &  $\log_{10}{f_{0,3}}$ & $\log_{10}{w_{3}}$  & $\log_{10}{A_{\rm 4}}$ &  $\log_{10}{f_{0,4}}$ & $\log_{10}{w_{4}}$ &($H_{0}$, $H_{1}$) & $T_{\rm LRT}$ & $p_{_{\rm LRT}}$ \\
			\hline
			\multirow{5}{*}{BAT} & \multirow{5}{*}{$[2.07, 22.07]$} & \multirow{5}{*}{$8$}	 & \multirow{5}{*}{$[100, 300]$} & PL  & $2.03^{+0.05}_{-0.06}$ & -- & $1.59^{+0.08}_{-0.08}$ & -- & $0.42^{+0.02}_{-0.02}$ & -- & -- & -- & -- & -- & -- & -- & -- & -- & -- & -- & -- & -- & -- & --\\	
			& 	&   &  & BPL & $2.34^{+0.34}_{-0.18}$ & $1.58^{+0.18}_{-0.55}$ & $1.55^{+0.21}_{-0.60}$ & $-0.67^{+0.66}_{-0.44}$ & $0.43^{+0.02}_{-0.02}$ & -- & -- & -- & -- & -- & -- & -- & -- & -- & -- & -- & -- & (PL, BPL) & {0.36} & {$6.34 \times 10^{-1}$}\\ 
                &  & & & PL+1QPO & $2.01^{+0.06}_{-0.05}$ & -- & $1.63^{+0.09}_{-0.08}$ & -- & $0.43^{+0.02}_{-0.02}$ & -- & -- & -- & -- & -- & -- & -- & -- & -- & $0.95^{+0.25}_{-0.49}$ & $0.80^{+0.01}_{-0.01}$ & $-0.85^{+0.24}_{-0.61}$ & (PL, PL+1QPO) & $13.64$ & $4.94 \times 10^{-3}$\\
                &  & & & PL+2QPOs & $1.95^{+0.06}_{-0.06}$ & -- & $1.67^{+0.10}_{-0.09}$ & -- & $0.45^{+0.02}_{-0.02}$ & -- & -- & -- & $1.36^{+0.44}_{-0.29}$ & $0.51^{+0.01}_{-0.01}$ & $-1.33^{+0.37}_{-0.68}$ & -- & -- & -- & $1.08^{+0.20}_{-0.25}$ & $0.81^{+0.01}_{-0.01}$ & $-0.73^{+0.16}_{-0.29}$ & (PL, PL+2QPOs) & $33.19$ & $1.34 \times 10^{-5}$\\
			&  &  &  & PL+4QPOs & $1.93^{+0.06}_{-0.06}$ & -- & $1.70^{+0.10}_{-0.09}$ & -- & $0.46^{+0.02}_{-0.02}$ & $-0.36^{+1.11}_{-1.13}$ & $0.37^{+0.04}_{-0.04}$ & $-1.77^{+0.83}_{-0.85}$ & $1.36^{+0.42}_{-0.27}$ &  $0.51^{+0.01}_{-0.01}$ & $-1.33^{+0.35}_{-0.66}$ & $0.26^{+0.78}_{-0.51}$ & $0.63^{+0.03}_{-0.02}$ & $-1.65^{+0.70}_{-0.92}$ & $1.11^{+0.18}_{-0.23}$ & $0.81^{+0.01}_{-0.01}$ & $-0.71^{+0.14}_{-0.27}$ & (PL, PL+4QPOs) & $41.79$ & {$3.44 \times 10^{-5}$} \\
			\hline
			\multirow{5}{*}{Fermi,n7} & \multirow{5}{*}{$[2.07, 22.07]$} & \multirow{5}{*}{$8$}	 & \multirow{5}{*}{$[100, 300]$} & PL  & $2.47^{+0.05}_{-0.05}$ & -- & $1.69^{+0.07}_{-0.06}$ & -- & $0.16^{+0.03}_{-0.04}$ & -- & -- & -- & -- & -- & -- & -- & -- & -- & -- & -- & -- & -- & -- & --\\	
			&  &  &  & BPL & $2.78^{+0.45}_{-0.20}$ & $1.67^{+0.19}_{-0.77}$ & $1.64^{+0.21}_{-0.77}$ & $-0.64^{+0.61}_{-0.44}$ & $0.18^{+0.03}_{-0.03}$& -- & -- & -- & -- & -- & -- & -- & -- & -- & -- & -- & -- & (PL, BPL) & $1.61$ &{$3.28 \times 10^{-1}$}\\ 
                &  & & & PL+1QPO & $2.44^{+0.05}_{-0.05}$ & -- & $1.71^{+0.07}_{-0.06}$ & -- & $0.19^{+0.03}_{-0.03}$ & -- & -- & -- & -- & -- & -- & -- & -- & -- & $1.29^{+0.26}_{-0.23}$ & $0.83^{+0.01}_{-0.01}$ & $-1.04^{+0.24}_{-0.41}$ & (PL, PL+1QPO) & $22.13$ & $9.58 \times 10^{-5}$\\
                &  & & & PL+2QPOs & $2.36^{+0.06}_{-0.05}$ & -- & $1.68^{+0.07}_{-0.06}$ & -- & $0.20^{+0.03}_{-0.03}$ & -- & -- & -- & $1.76^{+0.23}_{-0.22}$ & $0.52^{+0.01}_{-0.01}$ & $-1.04^{+0.27}_{-0.39}$ & -- & -- & -- & $1.34^{+0.22}_{-0.21}$ & $0.83^{+0.01}_{-0.01}$ & $-0.99^{+0.25}_{-0.35}$ & (PL, PL+2QPOs) & $45.42$ & $1.07 \times 10^{-7}$\\
			&  &  &  & PL+4QPOs & $2.35^{+0.06}_{-0.05}$ & -- & $1.68^{+0.06}_{-0.06}$ & -- & $0.20^{+0.03}_{-0.03}$ & $-0.17^{+1.23}_{-1.25}$ & $0.37^{+0.04}_{-0.04}$ & $-1.77^{+0.84}_{-0.85}$ & $1.76^{+0.23}_{-0.22}$ &  $0.52^{+0.01}_{-0.01}$ & $-1.04^{+0.27}_{-0.39}$ & $-0.12^{+1.05}_{-1.26}$ & $0.65^{+0.04}_{-0.04}$ & $-1.61^{+0.78}_{-0.97}$ & $1.34^{+0.21}_{-0.20}$ & $0.83^{+0.01}_{-0.01}$ & $-0.98^{+0.24}_{-0.34}$ & (PL, PL+4QPOs) & $48.37$ & {$4.60 \times 10^{-6}$} \\
			\hline
			\multirow{5}{*}{BAT} & \multirow{5}{*}{$[2.07, 22.07]$} & \multirow{5}{*}{$4$}	 & \multirow{5}{*}{$[100, 300]$} & PL  & $2.10^{+0.06}_{-0.06}$ & -- & $1.76^{+0.08}_{-0.08}$ & -- & $0.56^{+0.01}_{-0.01}$& -- & -- & -- & -- & -- & -- & -- & -- & -- & -- & -- & -- & -- & -- & --\\	
			&  &  &  & BPL & $2.44^{+0.33}_{-0.17}$ & $1.76^{+0.33}_{-0.76}$ & $1.67^{+0.42}_{-0.70}$ & $-0.24^{+0.42}_{-0.65}$ & $0.56^{+0.01}_{-0.01}$& -- & -- & -- & -- & -- & -- & -- & -- & -- & -- & -- & -- & (PL, BPL) & $6.30$ &{$2.46 \times 10^{-2}$}\\ 
                &  & & & PL+1QPO & $2.07^{+0.06}_{-0.06}$ & -- & $1.83^{+0.09}_{-0.09}$ & -- & $0.56^{+0.01}_{-0.01}$ & -- & -- & -- & -- & -- & -- & -- & -- & -- & $1.12^{+0.19}_{-0.23}$ & $0.81^{+0.01}_{-0.01}$ & $-0.73^{+0.16}_{-0.28}$ & (PL, PL+1QPO) & $18.63$ & $4.87 \times 10^{-4}$\\
                &  & & & PL+2QPOs & $1.96^{+0.06}_{-0.06}$ & -- & $1.90^{+0.11}_{-0.10}$ & -- & $0.57^{+0.01}_{-0.01}$ & -- & -- & -- & $1.47^{+0.28}_{-0.21}$ & $0.51^{+0.01}_{-0.01}$ & $-1.15^{+0.33}_{-0.48}$ & -- & -- & -- & $1.23^{+0.15}_{-0.18}$ & $0.81^{+0.01}_{-0.01}$ & $-0.66^{+0.12}_{-0.19}$ & (PL, PL+2QPOs) & $46.88$ & $5.91 \times 10^{-8}$\\
			&  &  &  & PL+4QPOs & $1.90^{+0.07}_{-0.08}$ & -- & $2.01^{+0.17}_{-0.13}$ & -- & $0.57^{+0.01}_{-0.01}$ & $0.63^{+0.68}_{-1.67}$ & $0.37^{+0.02}_{-0.03}$ & $-1.59^{+0.64}_{-0.93}$ & $1.48^{+0.33}_{-0.20}$ &  $0.51^{+0.01}_{-0.01}$ & $-1.22^{+0.29}_{-0.49}$ & $0.97^{+0.45}_{-0.44}$ & $0.63^{+0.01}_{-0.01}$ & $-1.41^{+0.44}_{-0.87}$ & $1.28^{+0.15}_{-0.15}$ & $0.81^{+0.01}_{-0.01}$ & $-0.63^{+0.09}_{-0.16}$ & (PL, PL+4QPOs) & $66.94$&{$1.58 \times 10^{-8}$}\\
			\hline
			\multirow{5}{*}{Fermi,n7} & \multirow{5}{*}{$[2.07, 22.07]$} & \multirow{5}{*}{$4$}	 & \multirow{5}{*}{$[100, 300]$} & PL  & $2.49^{+0.05}_{-0.05}$ & -- & $1.73^{+0.06}_{-0.06}$ & -- & $0.22^{+0.01}_{-0.01}$& -- & -- & -- & -- & -- & -- & -- & -- & -- & -- & -- & -- & -- & -- & --\\	
			&  &  &  & BPL & $2.80^{+0.41}_{-0.19}$ & $1.71^{+0.20}_{-0.77}$ & $1.68^{+0.23}_{-0.78}$ & $-0.50^{+0.58}_{-0.50}$ & $0.23^{+0.01}_{-0.01}$& -- & -- & -- & -- & -- & -- & -- & -- & -- & -- & -- & -- & (PL, BPL) & $4.07$&{$8.62 \times 10^{-2}$}\\ 
                &  & & & PL+1QPO & $2.46^{+0.05}_{-0.05}$ & -- & $1.74^{+0.06}_{-0.06}$ & -- & $0.22^{+0.01}_{-0.01}$ & -- & -- & -- & -- & -- & -- & -- & -- & -- & $1.32^{+0.22}_{-0.22}$ & $0.83^{+0.01}_{-0.01}$ & $-1.01^{+0.26}_{-0.38}$ & (PL, PL+1QPO) & $23.92$ & $4.17 \times 10^{-5}$\\
                &  & & & PL+2QPOs & $2.37^{+0.06}_{-0.05}$ & -- & $1.69^{+0.06}_{-0.05}$ & -- & $0.22^{+0.01}_{-0.01}$ & -- & -- & -- & $1.76^{+0.24}_{-0.22}$ & $0.52^{+0.01}_{-0.01}$ & $-1.04^{+0.28}_{-0.39}$ & -- & -- & -- & $1.36^{+0.21}_{-0.20}$ & $0.83^{+0.01}_{-0.01}$ & $-0.97^{+0.24}_{-0.33}$ & (PL, PL+2QPOs) & $46.15$ & $7.89 \times 10^{-8}$\\
			&  &  &  & PL+4QPOs & $2.36^{+0.05}_{-0.05}$ & -- & $1.69^{+0.06}_{-0.05}$ & -- & $0.23^{+0.01}_{-0.01}$ & $-0.18^{+1.23}_{-1.25}$ & $0.37^{+0.04}_{-0.04}$ & $-1.76^{+0.83}_{-0.87}$ & $1.76^{+0.24}_{-0.21}$ &  $0.52^{+0.01}_{-0.01}$ & $-1.05^{+0.28}_{-0.40}$ & $-0.10^{+1.06}_{-1.27}$ & $0.65^{+0.04}_{-0.04}$ & $-1.60^{+0.78}_{-0.96}$ & $1.36^{+0.21}_{-0.19}$ & $0.83^{+0.01}_{-0.01}$ & $-0.96^{+0.24}_{-0.32}$ & (PL, PL+4QPOs) & $50.42$ &{$2.46 \times 10^{-6}$}\\
			\hline
			\multirow{3}{*}{BAT} & \multirow{3}{*}{$[-10, 40]$} & \multirow{3}{*}{$16$}	 & \multirow{3}{*}{$[15, 30]$} & PL  & $1.70^{+0.04}_{-0.04}$ & -- & $2.60^{+0.11}_{-0.10}$ & -- & $0.39^{+0.01}_{-0.01}$ & -- & -- & -- & -- & -- & -- & -- & -- & -- & -- & -- & -- & -- & -- & --\\	
			&  &  &  & BPL & $2.97^{+0.43}_{-0.25}$ & $0.93^{+0.31}_{-0.37}$ & $3.11^{+0.18}_{-0.18}$ & $-0.52^{+0.09}_{-0.16}$ & $0.40^{+0.01}_{-0.01}$ & -- & -- & -- & -- & -- & -- & -- & -- & -- & -- & -- & -- & (PL, BPL) & $23.51$&{$4.12 \times 10^{-6}$}\\ 
			&  &  &  & BPL+3QPOs & $3.20^{+0.68}_{-0.60}$ & $0.82^{+0.54}_{-0.51}$ & $3.34^{+2.11}_{-0.43}$ & $-0.72^{+0.21}_{-0.24}$ & $0.40^{+0.01}_{-0.01}$ & $1.87^{+0.26}_{-0.41}$ & $-0.19^{+0.02}_{-0.03}$ & $-1.35^{+0.41}_{-0.49}$ & $1.50^{+0.16}_{-0.21}$ & $0.09^{+0.01}_{-0.02}$ & $-0.93^{+0.21}_{-0.24}$ &  $-1.01^{+1.20}_{-1.36}$ & $0.25^{+0.03}_{-0.03}$ & $-1.61^{+1.13}_{-0.92}$ & -- & -- & -- & (BPL, BPL+3QPOs) & {30.71}&{$5.93 \times 10^{-4}$}\\ 
			\hline
			\multirow{3}{*}{Fermi,n7} & \multirow{3}{*}{$[-10, 40]$} & \multirow{3}{*}{$16$}	 & \multirow{3}{*}{$[8, 30]$} & PL  & $1.22^{+0.04}_{-0.04}$ & -- & $2.42^{+0.11}_{-0.11}$ & -- & $0.28^{+0.01}_{-0.01}$ & -- & -- & -- & -- & -- & -- & -- & -- & -- & -- & -- & -- & -- & -- & --\\	
			&  &  &  & BPL & $2.71^{+0.78}_{-0.56}$ & $0.94^{+0.51}_{-0.59}$ & $2.74^{+0.24}_{-0.20}$ & $-0.81^{+0.26}_{-0.28}$ & $0.29^{+0.01}_{-0.01}$ & -- & -- & -- & -- & -- & -- & -- & -- & -- & -- & -- & -- & (PL, BPL) & {9.30}&{$7.18 \times 10^{-3}$}\\ 
			&  &  &  & BPL+3QPOs & $2.60^{+0.70}_{-0.56}$ & $0.93^{+0.53}_{-0.54}$ & $4.84^{+2.04}_{-1.34}$ & $-0.63^{+0.15}_{-0.21}$ & $0.29^{+0.01}_{-0.01}$ & $1.39^{+0.21}_{-0.21}$ & $-0.17^{+0.01}_{-0.02}$ & $-1.44^{+0.26}_{-0.34}$ & $0.96^{+0.12}_{-0.15}$ & $0.09^{+0.02}_{-0.02}$ & $-0.90^{+0.20}_{-0.22}$ &  $0.29^{+0.23}_{-0.27}$ & $0.28^{+0.01}_{-0.01}$ & $-1.40^{+0.34}_{-0.45}$ & -- & -- & -- & (BPL, BPL+3QPOs) & {39.60}&{$3.10 \times 10^{-5}$}\\ 
			\hline
			\multirow{3}{*}{BAT} & \multirow{3}{*}{$[-10, 40]$} & \multirow{3}{*}{$8$}	 & \multirow{3}{*}{$[15, 30]$} & PL  & $1.70^{+0.04}_{-0.04}$ & -- & $2.66^{+0.11}_{-0.10}$ & -- & $0.43^{+0.01}_{-0.01}$& -- & -- & -- & -- & -- & -- & -- & -- & -- & -- & -- & -- & -- & -- & --\\	
			&  &  &  & BPL & $2.97^{+0.34}_{-0.23}$ & $0.93^{+0.29}_{-0.32}$ & $3.21^{+0.17}_{-0.18}$ & $-0.49^{+0.07}_{-0.14}$ & $0.44^{+0.01}_{-0.01}$ & -- & -- & -- & -- & -- & -- & -- & -- & -- & -- & -- & -- & (PL, BPL) & {27.11}&{$6.21 \times 10^{-7}$}\\ 
			&  &  &  & BPL+3QPOs & $3.14^{+0.69}_{-0.53}$ & $0.85^{+0.49}_{-0.51}$ & $3.62^{+2.05}_{-0.56}$ & $-0.67^{+0.18}_{-0.22}$ & $0.44^{+0.01}_{-0.01}$ & $1.90^{+0.22}_{-0.35}$ & $-0.19^{+0.02}_{-0.02}$ & $-1.36^{+0.36}_{-0.47}$ & $1.53^{+0.14}_{-0.19}$ & $0.09^{+0.02}_{-0.02}$ & $-0.94^{+0.18}_{-0.24}$ &  $-0.81^{+1.14}_{-1.49}$ & $0.25^{+0.02}_{-0.03}$ & $-1.66^{+1.11}_{-0.87}$ & -- & -- & -- & (BPL, BPL+3QPOs) & {33.60}&{$2.27 \times 10^{-4}$}\\ 
			\hline
			\multirow{3}{*}{Fermi,n7} & \multirow{3}{*}{$[-10, 40]$} & \multirow{3}{*}{$8$}	 & \multirow{3}{*}{$[8, 30]$} & PL  & $1.22^{+0.04}_{-0.04}$ & -- & $2.44^{+0.11}_{-0.11}$ & -- & $0.28^{+0.01}_{-0.01}$& -- & -- & -- & -- & -- & -- & -- & -- & -- & -- & -- & -- & -- & -- & --\\	
			&  &  &  & BPL & $2.71^{+0.76}_{-0.56}$ & $0.92^{+0.51}_{-0.56}$ & $2.75^{+0.23}_{-0.19}$ & $-0.80^{+0.26}_{-0.27}$ & $0.29^{+0.01}_{-0.01}$ & -- & -- & -- & -- & -- & -- & -- & -- & -- & -- & -- & -- & (PL, BPL) & {9.71}&{$5.79 \times 10^{-3}$}\\ 
			&  &  &  & BPL+3QPOs & $2.61^{+0.70}_{-0.58}$ & $0.93^{+0.54}_{-0.54}$ & $4.84^{+2.08}_{-1.36}$ & $-0.63^{+0.15}_{-0.21}$ & $0.29^{+0.01}_{-0.01}$ & $1.40^{+0.21}_{-0.20}$ & $-0.17^{+0.01}_{-0.02}$ & $-1.44^{+0.26}_{-0.36}$ & $0.97^{+0.12}_{-0.13}$ & $0.09^{+0.02}_{-0.02}$ & $-0.89^{+0.19}_{-0.20}$ &  $0.29^{+0.24}_{-0.29}$ & $0.28^{+0.01}_{-0.01}$ & $-1.39^{+0.35}_{-0.46}$ & -- & -- & -- & (BPL, BPL+3QPOs) & {39.07}&{$3.70 \times 10^{-5}$}\\ 
			\hline
			\hline
		\end{tabular}
	}
    \begin{tablenotes}
\item All errors represent the $1 \sigma$ uncertainties.
\end{tablenotes}
\end{table*}

\clearpage
\begin{table*}
	\renewcommand\arraystretch{1.5}
	\centering
	\caption{Prior distributions for the Fourier periodogram fitting parameters.}
	\label{exttab4}
    \resizebox{\textwidth}{!}{
	\begin{tabular}{|c|c|c|c|}
		\hline
		\hline 
		Parameter & Meaning & Distribution for $100 \text{--} 300\,\mathrm{keV}$ & Distribution for $15 \text{--} 30\,\mathrm{keV}$\\
		\hline
	$\log_{10}{A}$ & Normalization factor for the PL or BPL model. & $\mathcal{U}(0, 7)$ &  $\mathcal{U}(0, 7)$  \\
	$\alpha$ & Low-frequency slope of the BPL model. & $\mathcal{U}(0, 2.5)$ &  $\mathcal{U}(0, 2.5)$ \\ 
	$\beta$ & High-frequency slope of the BPL model or the slope of the PL model. & $\mathcal{U}(0, 10)$ & $\mathcal{U}(1, 10)$ \\ 
	$\log_{10}{f_{\rm bend}}$ & Bend frequency of the BPL model & $\mathcal{U}(\log_{10}{0.05}, \log_{10}{2})$ & $\mathcal{U}(\log_{10}{0.02}, \log_{10}{0.4})$\\  
	$\log_{10}{C}$ & White noise constant for the PL or BPL models. & $\mathcal{U}(0, 1)$ & $\mathcal{U}(0, 1)$ \\ 
	$\log_{10}{A_{1}}$ & Normalization factor of the first QPO component. & $\mathcal{U}(-2, 3)$ & $\mathcal{U}(-2, 3)$ \\ 
	  $\log_{10}{f_{0,1}}$ & Central frequency of the first QPO component. & $\mathcal{U}(\log_{10}{2.0}, \log_{10}{2.8})$ & $\mathcal{U}(\log_{10}{0.4}, \log_{10}{0.9})$\\
	$\log_{10}{w_{1}}$ & Full Width at Half Maximum of the first QPO component. & $\mathcal{U}(-3, -0.5)$ & $\mathcal{U}(-3, 0)$\\ 
        $\log_{10}{A_{2}}$ & Normalization factor of the second QPO component. & $\mathcal{U}(-2, 3)$ & $\mathcal{U}(-2, 3)$ \\ 
	  $\log_{10}{f_{0,2}}$ & Central frequency of the second QPO component. & $\mathcal{U}(\log_{10}{2.8}, \log_{10}{3.7})$ & $\mathcal{U}(\log_{10}{0.9}, \log_{10}{1.6})$\\
	$\log_{10}{w_{2}}$ & Full Width at Half Maximum of the second QPO component. & $\mathcal{U}(-3, -0.5)$ & $\mathcal{U}(-3, 0)$\\
        $\log_{10}{A_{3}}$ & Normalization factor of the third QPO component. & $\mathcal{U}(-2, 3)$ & $\mathcal{U}(-2, 3)$ \\ 
	  $\log_{10}{f_{0,3}}$ & Central frequency of the third QPO component. & $\mathcal{U}(\log_{10}{3.7}, \log_{10}{5.2})$ & $\mathcal{U}(\log_{10}{1.6}, \log_{10}{2})$\\
	$\log_{10}{w_{3}}$ & Full Width at Half Maximum of the third QPO component. & $\mathcal{U}(-3, -0.5)$ & $\mathcal{U}(-3, 0)$\\
        $\log_{10}{A_{4}}$ & Normalization factor of the fourth QPO component. & $\mathcal{U}(-2, 3)$ & -- \\ 
	  $\log_{10}{f_{0,4}}$ & Central frequency of the fourth QPO component. & $\mathcal{U}(\log_{10}{5.2}, \log_{10}{7.5})$ & --\\
	$\log_{10}{w_{4}}$ & Full Width at Half Maximum of the fourth QPO component. & $\mathcal{U}(-3, -0.5)$ & --\\
		\hline
		\hline
	\end{tabular}
    }
\end{table*}

\clearpage
\begin{table*}
	\renewcommand\arraystretch{1.5}
	\centering
	\caption{The best-fit parameters of the periodic FRED pulses.}
	\label{exttab5}        
	\begin{tabular}{|c|c|c|}
		\hline
		\hline 
		Parameter & Prior & Posterior  \\
		$A_{1}$ & $\mathcal{U}(1, 300)$ & $112.27^{+68.80}_{-37.37}$ \\ 
		  $\tau$ & $\mathcal{U}(0.1, 1.5)$ & $0.87^{+0.05}_{-0.05}$ \\ 
	 	$\zeta$ & $\mathcal{U}(0.1, 3)$ & $1.79^{+0.24}_{-0.20}$ \\  
		$\Delta$ & $\mathcal{U}(0.5, 1.5)$ & $0.89^{+0.04}_{-0.04}$ \\ 
		$A_{2}$ & $\mathcal{U}(1, 300)$ & $64.21^{+41.41}_{-22.37}$ \\ 
		$A_{3}$ & $\mathcal{U}(1, 300)$ & $21.27^{+14.42}_{-7.76}$ \\
		$A_{4}$ & $\mathcal{U}(1, 300)$ & $17.05^{+10.91}_{-5.97}$ \\ 
        $\delta_{t}$ & $\mathcal{U}(1, 2)$ & $1.49^{+0.01}_{-0.01}$ \\
		\hline
		\hline
	\end{tabular}
    \begin{tablenotes}
\item All errors represent the $1 \sigma$ uncertainties.
\end{tablenotes}
\end{table*}

\clearpage
\bibliography{sn-bibliography}
\end{document}